\documentclass[prc,preprint,
  showpacs,showkeys,lengthcheck,
  nofootinbib,tightenlines,onecolumn,notitlepage,
  preprintnumbers,superscriptaddress]{revtex4-1}

\usepackage{graphicx}
\usepackage[usenames,dvipsnames]{xcolor}
\graphicspath{{figures/}}

\usepackage{hyperref}
\hypersetup{colorlinks=true,citecolor=blue,linkcolor=blue,urlcolor=blue}

\usepackage{diagbox}
\usepackage{booktabs}

\usepackage{amsmath,amssymb,amsfonts}
\usepackage{mathrsfs}
\usepackage{slashed}
\usepackage{bm}

\usepackage[draft]{changes} 

\newcommand{\olra}[1]{\overleftrightarrow{#1}}

\begin{document}

\title{A convolution formalism for defining densities of hadrons}

\author{Adam Freese}
\email{afreese@uw.edu}
\address{Department of Physics, University of Washington, Seattle, WA 98195, USA}

\author{Gerald A. Miller}
\email{miller@uw.edu}
\address{Department of Physics, University of Washington, Seattle, WA 98195, USA}

\begin{abstract}
  We clarify the meaning of spatial densities of hadrons.
  A physical density is given by the expectation value of a local operator
  for a physical state,
  and depends on both internal structure and the hadron's wave packet.
  In some particular cases, the physical density can be written as a convolution
  between a density function
  that depends on internal structure but not wave packet,
  and a smearing function
  that depends on wave packet but not internal structure.
  We show that the light front densities often encountered in the literature
  have this property
  but that instant form densities
  do not.
  For hadrons prepared in broad wave packets,
  physical instant form densities approximately obey such a convolution relation,
  with Breit frame densities as the apparent internal densities.
  However, there is an infinite series of relativistic corrections to
  this convolution formula.
\end{abstract}

\preprint{NT@UW-22-17}

\maketitle


\section{Introduction}
\label{sec:intro}

There has lately been renewed interest and debate about the proper manner
for describing the internal structure of hadrons in terms of spatial coordinates.
This debate is especially pertinent now with the increasing focus on the
energy momentum tensor and its associated densities in the
literature
(see e.g.\ Refs.~\cite{Polyakov:2018zvc,Lorce:2018egm,Freese:2021czn}).
A major focus of the upcoming Electron Ion Collider~\cite{Boer:2011fh,Accardi:2012qut,AbdulKhalek:2021gbh}
is to provide a spatial picture of the distributions of partons in hadrons,
making it vital that an accurate method for obtaining this spatial picture is used.

For a long time, authors have primarily calculated relativistic densities
through three-dimensional Fourier transforms of form factors,
giving a result referred to as the Breit frame density~\cite{Sachs:1962zzc}.
For almost as long, this approach has been criticized~\cite{Fleming:1974af,Burkardt:2000za,Miller:2018ybm}
for failing to follow from the elementary definition of a density
in relativistic field theory
(i.e., the expectation value of a local operator).
While the majority of authors continue to use the Breit frame density
uncritically,
multiple solutions to addressing the issue of relativistic spatial densities
have been advanced.

There are currently three established camps on the question of relativistic spatial densities.
The first camp uses the formalism of Wigner phase space distributions
to justify the Breit frame densities as a slice of a six-dimensional phase space distribution
when the hadron momentum is zero~\cite{Lorce:2018egm}.

The second camp obtains densities using expectation values of local currents
in light front coordinates, integrating out the coordinate
$x^-$ to produce a two-dimensional density in the transverse
plane~\cite{Burkardt:2002hr,Miller:2010nz,Freese:2021czn}.
Within this formalism, one often localizes the hadron's wave
packet~\cite{Burkardt:2000za,Freese:2021czn}
or works with Dirac delta functions~\cite{Diehl:2002he,Burkardt:2002hr,Miller:2018ybm}
to remove dependence on the wave packet.

The last camp attempts to define three-dimensional densities in
instant form coordinates in a similar manner to the second camp,
namely by taking the expectation value of a local current for a physical state
and localizing the wave packet.
This idea was first proposed by Fleming in 1974~\cite{Fleming:1974af},
but the idea has recently been revived~\cite{Epelbaum:2022fjc,Panteleeva:2022khw,Carlson:2022eps}.

Fleming's proposal was motivated by the fact that,
in non-relativistic quantum mechanics,
localizing the center-of-mass wave packet of a composite system
does reproduce the internal charge density of the system
as conventionally defined.
Without a clear notion of what constitutes the internal charge density
of a relativistic bound system,
using this procedure as a generalized \emph{definition} for the internal densities
seems reasonable.
However, the approach exhibits several shortcomings,
including the need to assume spherical symmetry of the wave packet during
localization~\cite{Fleming:1974af,Epelbaum:2022fjc,Panteleeva:2022khw,Carlson:2022eps,Panteleeva:2022uii,Panteleeva:2023evj},
suggesting lingering wave packet dependence.
Additionally, densities associated with the energy-momentum tensor
diverge~\cite{Panteleeva:2022uii,Panteleeva:2023evj}, a manifestation of the uncertainty principle.

The purpose of this work is to propose an alternative definition of internal densities
for relativistic composite systems,
which like Fleming's proposal reproduces known results when applied to
the non-relativistic domain.
Within this approach, we analyze when wave packet localization reproduces
the internal densities and when it does not.
In particular, we follow
a similar vein to recent work by Li \textsl{et al.}~\cite{Li:2022ldb},
with a focus on separating physical densities into
wave packet dependent and independent pieces.
Ultimately, we find that localization does reproduce our results
for a handful of light front densities,
but not for a class of cases we refer to as ``compound densities,''
which includes all densities in instant form coordinates.

This work is organized as follows.
In Sec.~\ref{sec:definitions},
we provide our definitions and the physical motivation behind them.
In Sec.~\ref{sec:lf},
we obtain light front densities within the constraints laid out in
the previous section.
These include the electromagnetic current and spatial distributions
associated with the energy-momentum tensor,
reproducing previous findings where they exist
but without requiring wave packet localization.
In Sec.~\ref{sec:3D},
we examine three-dimensional densities in instant form coordinates,
showing that exact three-dimensional internal densities do not exist
within our proposed formalism.
In the special case of diffuse wave packets
we find an infinite tower of internal densities that make increasingly
small contributions, with the Breit frame density appearing
as the leading term.
We summarize and conclude in Sec.~\ref{sec:conclusions}.


\section{Internal and physical densities}
\label{sec:definitions}

A physical density is defined as the expectation value of a local operator
for a physical state $|\Psi\rangle$ in Hilbert space:
\begin{align}
  \label{eqn:density:def}
  \rho_{\mathrm{phys}}(x)
  =
  \langle \Psi | \hat{O}(x) | \Psi \rangle
  \,.
\end{align}
We assume $|\Psi\rangle$ is a single-hadron state.
The physical density will contain influences both from the internal structure
of the hadron, and from the wave packet that it's prepared in.
The goal of hadron structure research is to isolate internal properties
of hadrons that are independent of artifacts of state preparation.
It is thus vital to determine under what circumstances a
purely internal hadron density can be defined.

It is moreover prudent to understand how the internal densities
contribute to the physical densities defined in Eq.~(\ref{eqn:density:def}).
The simplest scenario that can occur is that the physical density
corresponds to an internal density being smeared over space
at any particular fixed time $\tau$
(which could be time in any of the forms of relativistic
dynamics~\cite{Dirac:1949cp}):
\begin{align}
  \label{eqn:density:simple}
  \rho_{\mathrm{phys}}(\mathbf{x},\tau)
  =
  \int \mathrm{d}^n \mathbf{x} \,
  \mathcal{S}_\Psi(\mathbf{R},\tau)
  \rho_{\mathrm{int.}}(\mathbf{x}-\mathbf{R})
  \,.
\end{align}
Here $\mathcal{S}_\Psi(\mathbf{R},\tau)$ is a smearing function that depends on
the wave packet but not on internal structure,
whereas $\rho_{\mathrm{int.}}(\mathbf{b})$ contains no wave packet dependence.
We have notated $n$ spatial dimensions for generality;
we will have $n=3$ for non-relativistic densities and $n=2$ for
relativistic light front densities.
We shall refer to the density in a scenario such as
Eq.~(\ref{eqn:density:simple}) as a \emph{simple density}.

Eq.~(\ref{eqn:density:simple}) has a clear physical interpretation:
$\rho_{\mathrm{int.}}(\mathbf{b})$
describes the distribution of some quantity
(such as charge or energy)
a displacement $\mathbf{b}$ from the
barycentric position $\mathbf{R}$
of the hadron,
and $\mathcal{S}_\Psi(\mathbf{R},\tau)$ gives a distribution
for the hadron's barycenter itself.
The convolution of both then gives the physical density
at a position $\mathbf{x} = \mathbf{R} + \mathbf{b}$.
We shall show in Sec.~\ref{sec:ex:nr} how such a formula naturally arises
in the context of non-relativistic quantum mechanics.

In an old work that has found recent attention,
Fleming~\cite{Fleming:1974af} noticed that the non-relativistic
charge density took the form of Eq.~(\ref{eqn:density:simple}),
with the smearing function being the probability density
$S_\Psi(\mathbf{R}) = \Psi^*(\mathbf{R}) \Psi(\mathbf{R})$.
If the wave packet is localized in this case,
the physical density approaches the internal density.
In fact, this will always happen when the physical density is a simple density
and the smearing function is the barycentric probability density.
Fleming extrapolated beyond this simple case and postulated that
localizing the wave function would always produce an internal density.

However, localization will not reproduce the desired internal densities
whenever the physical density does not obey Eq.~(\ref{eqn:density:simple}).
There will be cases that the physical density is actually
a sum of different internal structures,
each weighted by a different smearing function:
\begin{align}
  \label{eqn:density:compound}
  \rho_{\mathrm{phys}}(\mathbf{x},\tau)
  =
  \int \mathrm{d}^n \mathbf{x} \,
  \mathcal{S}_\Psi^{(1)}(\mathbf{R},\tau)
  \rho_{\mathrm{int.}}^{(1)}(\mathbf{x}-\mathbf{R},\tau)
  +
  \int \mathrm{d}^n \mathbf{x} \,
  \mathcal{S}_\Psi^{(2)}(\mathbf{R},\tau)
  \rho_{\mathrm{int.}}^{(2)}(\mathbf{x}-\mathbf{R},\tau)
  +
  \ldots
\end{align}
Here,
$\rho_{\mathrm{int.}}^{(1)}(\mathbf{x}-\mathbf{R},\tau)$
and
$\rho_{\mathrm{int.}}^{(2)}(\mathbf{x}-\mathbf{R},\tau)$
correspond to different internal spatial distributions,
which are each independent of the wave packet $\Psi$.
We shall refer to such a density as a \emph{compound density}.
In the light front formalism (Sec.~\ref{sec:lf}),
we shall see that the energy density and stress tensor
are examples of compound densities.
We shall also see---assuming convergence of the series---that
instant form densities are always compound densities with
infinitely many contributions,
as previously found by Li \textsl{et al.} in Ref.~\cite{Li:2022ldb}.

\subsection{Example from non-relativistic quantum mechanics}
\label{sec:ex:nr}

Let us consider a simple non-relativistic example of where
Eq.~(\ref{eqn:density:simple}) naturally arises.
We consider two particles with masses
$m_1$ and $m_2$ bound to a system with mass $M = m_1+m_2-\varepsilon$.
The wave function for the barycentric position is $\Psi(\mathbf{R},t)$
and the wave function for the relative position is
$\psi(\mathbf{r}) e^{i\varepsilon t}$.
If the positions of the particles are $\mathbf{r}_1$ and $\mathbf{r}_2$,
the barycentric and relative coordinates are defined:
\begin{subequations}
  \label{eqn:Rr}
  \begin{align}
    \mathbf{R}
    &=
    \frac{m_1 \mathbf{r}_1 + m_2 \mathbf{r}_2}{m_1+m_2}
    \\
    \mathbf{r}
    &=
    \mathbf{r}_2 - \mathbf{r}_1
    \,.
  \end{align}
\end{subequations}
We define a density for some quantity $Q$,
which may be electric charge or mass for instance,
and for which particles 1 and 2 carry an amount
$q_1$ and $q_2$ respectively.
The density is:
\begin{align}
  \label{eqn:rhoQ}
  \rho_Q(\mathbf{x},t)
  =
  \int \mathrm{d}^3 \mathbf{R}
  \int \mathrm{d}^3 \mathbf{r}
  \,
  |\Psi(\mathbf{R},t)|^2
  |\psi(\mathbf{r})|^2
  \Big\{
    q_1 \delta^{(3)}(\mathbf{x} - \mathbf{r}_1)
    +
    q_2 \delta^{(3)}(\mathbf{x} - \mathbf{r}_2)
    \Big\}
  \,.
\end{align}
To integrate out the delta functions, we must use
the inversion of Eq.~(\ref{eqn:Rr}), which gives:
\begin{align}
  \rho_Q(\mathbf{x},t)
  =
  \int \mathrm{d}^3 \mathbf{R}
  \,
  |\Psi(\mathbf{R},t)|^2
  \left\{
    q_1
    \left|\psi\left(\frac{m_1+m_2}{m_2}(\mathbf{R}-\mathbf{x})\right)\right|^2
    +
    q_2
    \left|\psi\left(\frac{m_1+m_2}{m_1}(\mathbf{x}-\mathbf{R})\right)\right|^2
    \right\}
  \,.
\end{align}
This has the form of Eq.~(\ref{eqn:density:simple}), with:
\begin{subequations}
  \begin{align}
    \mathcal{S}_\Psi(\mathbf{R},t)
    &=
    |\Psi(\mathbf{R},t)|^2
    \\
    \rho_{\mathrm{int.}}(\mathbf{b})
    &=
    q_1
    \left|\psi\left(-\frac{m_1+m_2}{m_2}\,\mathbf{b}\right)\right|^2
    +
    q_2
    \left|\psi\left(\frac{m_1+m_2}{m_1}\,\mathbf{b}\right)\right|^2
    \,.
  \end{align}
\end{subequations}
If, at some moment of time $t_0$ we have
$|\Psi(\mathbf{R},t_0)|^2 = \delta^{3}(\mathbf{R})$,
then we obtain
$\rho_Q(\mathbf{x},t_0) = \rho_{\mathrm{int.}}(\mathbf{x})$.
This is a generic feature of \emph{simple} densities where the smearing
function is equal to the probability distribution.


\section{Internal densities in light front coordinates}
\label{sec:lf}

In this section, we demonstrate that the light front densities encountered in
the literature~\cite{Burkardt:2000za,Burkardt:2002hr,Miller:2007uy,Carlson:2007xd,Carlson:2008zc,Miller:2009qu,Lorce:2018egm,Freese:2021czn,Freese:2022yur,Freese:2022ibw}
are faithfully reproduced in our formalism.
We also determine how they relate to their associated physical densities.
Although much of the previous literature tended to use a localization procedure
with a specific wave packet form
(or simply used delta function wave packets)
to obtain these results,
we will show that the results are in fact independent of the wave packet.

For simplicity, we will focus exclusively on spin-zero targets.
This work is meant to provide conceptual clarity and rigor to the
research program of hadron densities rather than to be an exhaustive catalogue.
Higher-spin targets with general polarization will contain complications from
helicity-flip contributions~\cite{Carlson:2007xd,Carlson:2008zc,Freese:2021mzg,Freese:2022yur},
so densities that are simple for spin-zero targets will often be
compound for higher-spin targets.


\subsection{General expression for physical densities}

We will derive general expression for the \emph{physical} light front densities
of spin-zero targets, without making any specific assumptions about
the wave packet.

The Lorentz-invariant completeness relation for states in the Hilbert
subspace of single-particle spin-zero states is:
\begin{align}
  \label{eqn:completeness}
  \int \frac{\mathrm{d}p^+ \mathrm{d}^2\mathbf{p}_\perp}{2p^+(2\pi)^3}
  | p^+, \mathbf{p}_\perp \rangle
  \langle p^+, \mathbf{p}_\perp |
  =
  1
  \,.
\end{align}
We define the momentum space wave function as:
\begin{subequations}
  \begin{align}
    \label{eqn:wf:p}
    \Psi(p^+,\mathbf{p}_\perp)
    =
    \frac{1}{\sqrt{2p^+}}
    \langle p^+, \mathbf{p}_\perp | \Psi \rangle
    \,,
  \end{align}
  so that it obeys the normalization condition:
  \begin{align}
    \int \frac{\mathrm{d}p^+ \mathrm{d}^2\mathbf{p}_\perp}{(2\pi)^3}
    \Big|
    \Psi(p^+,\mathbf{p}_\perp)
    \Big|^2
    =
    1
    \,.
  \end{align}
\end{subequations}
We define the position-space wave function through a Fourier transform:
\begin{subequations}
  \label{eqn:fourier}
  \begin{align}
    \Psi(z^-,\mathbf{z}_\perp)
    &=
    \int \frac{\mathrm{d}p^+ \mathrm{d}^2\mathbf{p}_\perp}{(2\pi)^3}
    \Psi(p^+,\mathbf{p}_\perp)
    e^{-ip^+z^-}
    e^{+i\mathbf{p}_\perp\cdot\mathbf{z}_\perp}
    \,,
  \end{align}
  so that it obeys the normalization condition:
  \begin{align}
    \int \mathrm{d}z^-
    \int \mathrm{d}^2\mathbf{z}_\perp \,
    \Big|
    \Psi(z^-,\mathbf{z}_\perp)
    \Big|^2
    =
    1
    \,.
  \end{align}
\end{subequations}
Using the completeness relation twice
with momenta $p$ and $p'$, integrating out $x^-$,
and defining a change of variables:
\begin{subequations}
  \begin{align}
    \mathbf{P}_\perp
    &=
    \frac{1}{2} \Big( \mathbf{p}_\perp + \mathbf{p}'_\perp \Big)
    \\
    \boldsymbol{\Delta}_\perp
    &=
    \mathbf{p}'_\perp - \mathbf{p}_\perp
    \\
    P^+
    &=
    p^+ = p'^+
  \end{align}
\end{subequations}
we obtain:
\begin{align}
  \rho_{\mathrm{LF}}(\mathbf{x}_\perp)
  =
  \int \frac{\mathrm{d}P^+\mathrm{d}^2\mathbf{P}_\perp}{2P^+(2\pi)^3}
  \int \frac{\mathrm{d}^2\boldsymbol{\Delta}_\perp}{(2\pi)^2}
  \langle \Psi | P^+, \mathbf{p}'_\perp \rangle
  \frac{
    \langle P^+,\mathbf{p}'_\perp | \hat{O}(0) | P^+,\mathbf{p}_\perp \rangle
  }{2P^+}
  \langle P^+, \mathbf{p}_\perp | \Psi\rangle
  e^{-i\boldsymbol{\Delta}_\perp \cdot \mathbf{x}_\perp}
  \,.
\end{align}
Rewriting in terms of the position-space wave functions gives:
\begin{multline}
  \rho_{\mathrm{LF}}(\mathbf{x}_\perp)
  =
  \int \mathrm{d}R'^-
  \int \mathrm{d}^2\mathbf{R}'_\perp
  \int \mathrm{d}R^-
  \int \mathrm{d}^2\mathbf{R}_\perp
  \int \frac{\mathrm{d}P^+\mathrm{d}^2\mathbf{P}_\perp}{(2\pi)^3}
  \int \frac{\mathrm{d}^2\boldsymbol{\Delta}_\perp}{(2\pi)^2}
  e^{-iP^+(R'^--R^-)}
  e^{i\mathbf{P}_\perp\cdot(\mathbf{R}'_\perp-\mathbf{R}_\perp)}
  \\
  e^{-i\boldsymbol{\Delta}_\perp \cdot \left(\mathbf{x}_\perp - \frac{\mathbf{R}_\perp+\mathbf{R}'_\perp}{2}\right)}
  \Psi^*(R'^-,\mathbf{R}'_\perp)
  \frac{
    \langle P^+,\mathbf{p}'_\perp | \hat{O}(0) | P^+,\mathbf{p}_\perp \rangle
  }{2P^+}
  \Psi(R^-,\mathbf{R}_\perp)
  \,.
\end{multline}
To proceed, the dependence on $P^+$ and $\mathbf{P}_\perp$
must be removed from the matrix element.
This can be accomplished by noting (for instance):
\begin{align*}
  \frac{i}{2}
  \Big( \nabla_i^{(R)} - \nabla_i^{(R')} \Big)
  e^{i\mathbf{P}_\perp\cdot(\mathbf{R}'_\perp-\mathbf{R}_\perp)}
  &=
  P_\perp^i
  e^{i\mathbf{P}_\perp\cdot(\mathbf{R}'_\perp-\mathbf{R}_\perp)}
  \\
  \frac{i}{2}
  \Big( \nabla_i^{(R)} - \nabla_i^{(R')} \Big)
  e^{-i\boldsymbol{\Delta}_\perp \cdot \left(\mathbf{x}_\perp - \frac{\mathbf{R}_\perp+\mathbf{R}'_\perp}{2}\right)}
  &=
  0
  \,.
\end{align*}
Thus any factors of $P_\perp^i$ that appear in the matrix element
can be transformed into differences of gradients,
and integration by parts can be used to move the gradients
to act on the barycentric wave functions.
A similar trick can be applied to $P^+$,
with the end result being that the following substitutions can be made:
\begin{subequations}
  \label{eqn:sub}
  \begin{align}
    \label{eqn:sub:perp}
    P_\perp^i
    \rightarrow
    -
    \frac{i}{2}
    \olra{\nabla}_i
    \\
    P^+
    \rightarrow
    +
    \frac{i}{2}
    \olra{\partial}_-
    \,,
  \end{align}
\end{subequations}
with the two-sided derivatives to be placed between the barycenter
wave function and its conjugate.
With these substitutions made,
it is possible to do the $P^+$ and $\mathbf{P}_\perp$ integrals,
which produce delta functions, and these delta functions can be
eliminated by doing the $R'^-$ and $\mathbf{R}'_\perp$ integrals.
Performing these steps gives:
\begin{align}
  \label{eqn:density:general}
  \rho_{\mathrm{LF}}(\mathbf{x}_\perp)
  =
  \int \mathrm{d}R^-
  \int \mathrm{d}^2\mathbf{R}_\perp
  \int \frac{\mathrm{d}^2\boldsymbol{\Delta}_\perp}{(2\pi)^2}
  e^{-i\boldsymbol{\Delta}_\perp \cdot \left(\mathbf{x}_\perp - \mathbf{R}_\perp\right)}
  \Psi^*(R^-,\mathbf{R}_\perp)
  \frac{
    \langle P^+,\mathbf{p}'_\perp | \hat{O}(0) | P^+,\mathbf{p}_\perp \rangle
  }{2P^+}
  \Psi(R^-,\mathbf{R}_\perp)
  \,,
\end{align}
where any $P^+$ or $P_\perp^i$ appearing in this formula should be
understood in terms on the substitutions of Eq.~(\ref{eqn:sub}).
To proceed any further, specific local operators must be considered.


\subsection{Electromagnetic current}

Let us first consider electromagnetic current of a spin-zero hadron.
The matrix element of the electromagnetic four-current $j^\mu(0)$
between momentum kets is:
\begin{align}
  \langle P^+, \mathbf{p}'_\perp |
  j^\mu(0)
  | P^+, \mathbf{p}_\perp \rangle
  =
  2 P^\mu F(t)
  \,,
\end{align}
where $F(t)$ is the electromagnetic form factor, and:
\begin{align}
  \label{eqn:t}
  t
  =
  (p'-p)^2
  =
  -\boldsymbol{\Delta}_\perp^2
  \,.
\end{align}
For the plus component of this current---the light front charge density---the
physical density takes a simple form:
\begin{align}
  \label{eqn:j+:phys}
  j^+_{\mathrm{LF}}(\mathbf{x}_\perp)
  =
  \int \mathrm{d}R^-
  \int \mathrm{d}^2\mathbf{R}_\perp \,
  \big| \Psi(R^-,\mathbf{R}_\perp) \big|^2
  \int \frac{\mathrm{d}^2\boldsymbol{\Delta}_\perp}{(2\pi)^2}
  F(-\boldsymbol{\Delta}_\perp^2)
  e^{-i\boldsymbol{\Delta}_\perp \cdot (\mathbf{x}_\perp - \mathbf{R}_\perp)}
  \,.
\end{align}
This constitutes a simple density in the form of Eq.~(\ref{eqn:density:simple}),
with an internal density:
\begin{align}
  \label{eqn:j+:true}
  j^+_{\mathrm{int.}}(\mathbf{b}_\perp)
  &=
  \int \frac{\mathrm{d}^2\boldsymbol{\Delta}_\perp}{(2\pi)^2}
  F(-\boldsymbol{\Delta}_\perp^2)
  \,
  e^{-i\boldsymbol{\Delta}_\perp \cdot \mathbf{b}_\perp}
  \,,
\end{align}
which is the standard result~\cite{Miller:2010nz},
and a smearing function:
\begin{align}
  \label{eqn:smear:P}
  \mathcal{P}(\mathbf{R}_\perp)
  &=
  \int \mathrm{d}R^- \,
  \big| \Psi(R^-,\mathbf{R}_\perp) \big|^2
  \,,
\end{align}
which is the transverse probability density for the barycentric position.
We have thus obtained the standard result for the light front charge density
of a spin-zero target, but without making any assumptions about the hadron
wave packet nor localizing it.
Note, however, that if
$\mathcal{P}(\mathbf{R}_\perp) = \delta^{(2)}(\mathbf{R}_\perp)$,
then
$j^+_{\mathrm{LF}}(\mathbf{x}_\perp) = j^+_{\mathrm{int.}}(\mathbf{x}_\perp)$,
which is why previous treatments with localized states obtained
the correct internal charge density.

It is possible to also obtain the transverse current.
One need only use $\mu=1,2$ instead of $\mu=+$.
It will be necessary to use the rules of Eq.~(\ref{eqn:sub}):
\begin{align}
  \frac{
    \langle P^+, \mathbf{p}'_\perp |
    j_\perp^i(0)
    | P^+, \mathbf{p}_\perp \rangle
  }{2P^+}
  =
  \frac{P_\perp^i}{P^+}
  F(t)
  \rightarrow
  -
  \frac{\olra{\nabla}_i}{\olra{\partial}_-}
  F(t)
  \,.
\end{align}
Since $R^-$ is integrated out in defining smearing functions,
it is possible to turn the $\olra{\partial}_-$ into an expectation value:
\begin{align}
  -
  \frac{\olra{\nabla}_i}{\olra{\partial}_-}
  F(t)
  \rightarrow
  \frac{ - i \olra{\nabla}_i }{2}
  \left\langle\frac{1}{P^+}\right\rangle
  F(t)
  \,.
\end{align}
We find thus that:
\begin{align}
  \mathbf{j}_\perp(\mathbf{x}_\perp)
  &=
  \left\langle\frac{1}{P^+}\right\rangle
  \int \mathrm{d}^2\mathbf{R}_\perp \,
  \boldsymbol{\mathcal{V}}_\perp(\mathbf{R}_\perp)
  \,
  j^+_{\mathrm{int.}}(\mathbf{x}_\perp-\mathbf{R}_\perp)
  \\
  \label{eqn:smear:V}
  \boldsymbol{\mathcal{V}}_\perp(\mathbf{R}_\perp)
  &=
  \int \mathrm{d}R^- \,
  \Psi^*(R^-,\mathbf{R}_\perp)
  \frac{-i\olra{\boldsymbol{\nabla}}_\perp}{2}
  \Psi(R^-,\mathbf{R}_\perp)
  \,.
\end{align}
The transverse electromagnetic current is thus also a simple density,
and in fact involves the same internal density as the electric charge density.
The only difference is how the smearing function distributes the internal density.


\subsection{Momentum densities and energy density}

We next consider momentum and energy densities,
which are given by matrix elements of the energy-momentum tensor (EMT).
Matrix elements between momentum kets are~\cite{Polyakov:2018zvc}:
\begin{align}
  \langle P^+, \mathbf{p}'_\perp |
  T^{\mu\nu}(0)
  | P^+, \mathbf{p}_\perp \rangle
  =
  2 P^\mu P^\nu A(t)
  +
  \frac{\Delta^\mu \Delta^\nu - g^{\mu\nu} \Delta^2}{2}
  D(t)
  \,,
\end{align}
where $A(t)$ and $D(t)$ are called gravitational form factors~\cite{Kobzarev:1962wt,Pagels:1966zza,Polyakov:2018zvc},
and $t$ is given in Eq.~(\ref{eqn:t}).
The momentum and energy densities are given by considering $\mu=+$ specifically,
for which:
\begin{align}
  \frac{
    \langle P^+, \mathbf{p}'_\perp |
    T^{+\nu}(0)
    | P^+, \mathbf{p}_\perp \rangle
  }{2P^+}
  =
  P^\nu A(t)
  +
  g^{+\nu}
  \frac{\boldsymbol{\Delta}_\perp^2}{4P^+}
  D(t)
  \,.
\end{align}
The momenta densities are those with $\nu=+,1,2$,
while the energy density is given by $\nu=-$
and is the only density to which $D(t)$ contributes.

The $P^+$ density is the most straightforward,
since the $P^+$ multiplying $A(t)$ just becomes an expectation value.
The physical density is given by:
\begin{align}
  T^{++}_{\mathrm{LF}}(\mathbf{x}_\perp)
  =
  \langle P^+ \rangle
  \int \mathrm{d}^2\mathbf{R}_\perp \,
  \mathcal{P}(\mathbf{R}_\perp)
  T^{++}_{\mathrm{int.}}(\mathbf{x}_\perp - \mathbf{R}_\perp)
  \,,
\end{align}
where the smearing function is defined in Eq.~(\ref{eqn:smear:P})
and the internal density is given by:
\begin{align}
  T^{++}_{\mathrm{int.}}(\mathbf{b}_\perp)
  =
  \int \frac{\mathrm{d}^2\boldsymbol{\Delta}_\perp}{(2\pi)^2}
  A(-\boldsymbol{\Delta}_\perp^2)
  \,
  e^{-i\boldsymbol{\Delta}_\perp\cdot\mathbf{b}_\perp}
  \,,
\end{align}
which is the standard result~\cite{Lorce:2018egm,Freese:2021czn}.
Although it was unnecessary to consider a localized wave packet to
obtain this result,
if we consider a localized state
$\mathcal{P}(\mathbf{R}_\perp) = \delta^{(2)}(\mathbf{R}_\perp)$,
then
$T^{++}_{\mathrm{LF}}(\mathbf{x}_\perp) = \langle P^+\rangle T^{++}_{\mathrm{int.}}(\mathbf{x}_\perp)$,
which is why Ref.~\cite{Freese:2021czn}
obtained the correct result.

The $\mathbf{P}_\perp$ density is similarly straightforward.
The substitution rule of Eq.~(\ref{eqn:sub:perp}) must be used,
which ends up giving:
\begin{align}
  \mathbf{T}^{+i}_{\mathrm{LF}}(\mathbf{x}_\perp)
  =
  \int \mathrm{d}^2\mathbf{R}_\perp \,
  \boldsymbol{\mathcal{V}}_\perp^i(\mathbf{R}_\perp)
  T^{++}_{\mathrm{int.}}(\mathbf{x}_\perp - \mathbf{R}_\perp)
  \,,
\end{align}
where the smearing function was defined in Eq.~(\ref{eqn:smear:V}).
So far, the situation is similar to what we saw for the electromagnetic current:
the intrinsic $P^+$ density can be convolved with a different smearing function
to obtain either the physical $P^+$ density or the physical $\mathbf{P}_\perp$
density.
The former is in fact proportional to the barycentric probability density.

Let us look at the energy density (i.e., the $P^-$ density) next.
The matrix element contains a factor $P^-$, which is given by:
\begin{align}
  P^-
  \equiv
  \frac{p^- + p'^-}{2}
  =
  \frac{M^2 + \mathbf{P}_\perp^2 + \frac{1}{4}\boldsymbol{\Delta}_\perp^2}{2P^+}
  \,.
\end{align}
The appropriate matrix element thus takes the form:
\begin{multline}
  \frac{
    \langle P^+, \mathbf{p}'_\perp |
    T^{+-}(0)
    | P^+, \mathbf{p}_\perp \rangle
  }{2P^+}
  =
  \frac{M^2 + \mathbf{P}_\perp^2 + \frac{1}{4}\boldsymbol{\Delta}_\perp^2}{2P^+}
  A(t)
  +
  \frac{\boldsymbol{\Delta}_\perp^2}{4P^+}
  D(t)
  \\
  \rightarrow
  \left\langle{\frac{1}{2P^+}}\right\rangle
  \left(
  \left(
    M^2
    - \frac{1}{4}\olra{\boldsymbol{\nabla}}_\perp^2
    + \frac{1}{4}\boldsymbol{\Delta}_\perp^2
    \right)
  A(t)
  +
  \frac{\boldsymbol{\Delta}_\perp^2}{2}
  D(t)
  \right)
  \,,
\end{multline}
where we used the substitution rule in Eq.~(\ref{eqn:sub:perp}).
To proceed, we need to define both a new smearing function:
\begin{align}
  \label{eqn:smear:W:-}
  \mathcal{W}^-(\mathbf{R}_\perp)
  &=
  \int \mathrm{d}R^- \,
  \Psi^*(R^-,\mathbf{R}_\perp)
  \frac{-\olra{\boldsymbol{\nabla}}_\perp^2}{4}
  \Psi(R^-,\mathbf{R}_\perp)
  \,,
\end{align}
and a new internal density:
\begin{align}
  T^{+-}_{\mathrm{int.}}(\mathbf{b}_\perp)
  =
  \int \frac{\mathrm{d}^2\boldsymbol{\Delta}_\perp}{(2\pi)^2}
  \left(
  \left(
    M^2
    +
    \frac{\boldsymbol{\Delta}_\perp^2}{4}
    \right)
  A(-\boldsymbol{\Delta}_\perp^2)
  +
  \frac{\boldsymbol{\Delta}_\perp^2}{2}
  D(-\boldsymbol{\Delta}_\perp^2)
  \right)
  e^{-i\boldsymbol{\Delta}_\perp\cdot\mathbf{b}_\perp}
  \,.
\end{align}
With this, we find the physical energy density to be a compound density:
\begin{align}
  \label{eqn:density:energy}
  T^{+-}_{\mathrm{LF}}(\mathbf{x}_\perp)
  &=
  \left\langle\frac{1}{2P^+}\right\rangle
  \left(
  \int \mathrm{d}^2\mathbf{R}_\perp \,
  \mathcal{W}^-(\mathbf{R}_\perp)
  T^{++}_{\mathrm{int.}}(\mathbf{x}_\perp-\mathbf{R}_\perp)
  +
  \int \mathrm{d}^2\mathbf{R}_\perp \,
  \mathcal{P}(\mathbf{R}_\perp)
  T^{+-}_{\mathrm{int.}}(\mathbf{x}_\perp-\mathbf{R}_\perp)
  \right)
  \,.
\end{align}
This compound density exhibits a curious structure.
The first term can be interpreted as a kinetic term
and the second as a dynamical term.
The smearing function $\mathcal{W}^-(\mathbf{R}_\perp)$ in the first term
arises from $\mathbf{P}_\perp^2$ in the momentum-space matrix element,
and describes the kinetic motion of the barycenter.
The convolution with the internal $P^+$ density then distributes
the barycentric kinetic motion over the constituents.
Because of this first term, if
$\mathcal{P}(\mathbf{R}_\perp) = \delta^{(2)}(\mathbf{R}_\perp)$,
then $T^{+-}_{\mathrm{LF}}(\mathbf{x}_\perp)$ diverges.
This is in effect a manifestation of the
uncertainty principle~\cite{Miller:2018ybm}.

The second term in Eq.~(\ref{eqn:density:energy}) encodes every
contribution to the hadron energy aside from barycentric kinetic motion.
In models of internal hadron dynamics that frame the hadron as a many-body system,
this includes masses of constituents,
kinetic energy of the constituents relative to the center-of-momentum,
and potential energy.
This can be seen by considering the general form of the
light-front Hamiltonian for an $N$-body system:
\begin{align}
  H
  =
  \sum_{i=1}^N
  \frac{m_i^2 + \mathbf{p}_{i\perp}^2}{2p_i^+}
  +
  V
  \,.
\end{align}
Here, $V$ is the potential energy,
and the absolute momenta are related to relative momenta
and light front momentum fractions by:
\begin{subequations}
  \begin{align}
    p_i^+
    &=
    x_i P^+
    \\
    \mathbf{p}_{i\perp}
    &=
    x_i \mathbf{P}_\perp
    +
    \mathbf{k}_{i\perp}
    \,,
  \end{align}
  and these obey sum rules:
  \begin{align}
    \sum_{i=1}^N
    x_i
    &=
    1
    \\
    \sum_{i=1}^N
    \mathbf{k}_{i\perp}
    &=
    0
    \,.
  \end{align}
\end{subequations}
Using these definitions and the sum rules, it follows:
\begin{align}
  \label{eqn:H:N}
  H
  =
  \frac{\mathbf{P}_\perp^2}{2P^+}
  +
  \frac{1}{2P^+}
  \sum_{i=1}^N
  \frac{m_i^2 + \mathbf{k}_{i\perp}^2}{x_i}
  +
  V
  \,.
\end{align}
The $\mathbf{P}_\perp^2/(2P^+)$ term of course corresponds to the first term
in Eq.~(\ref{eqn:density:energy}),
and thus the remaining terms in Eq.~(\ref{eqn:H:N})---which are exactly the aspects of the internal
energy we have listed---must correspond to the remaining second term of Eq.~(\ref{eqn:density:energy}).
It is for this reason that we identify
$T^{+-}_{\mathrm{int.}}(\mathbf{b}_\perp)$
as the \emph{internal} energy density.
In fact, the presence of $D(t)$ in this density, which has a well-established
connection to internal stresses~\cite{Polyakov:2018zvc,Polyakov:2018rew,Lorce:2018egm,Freese:2021czn},
helps indicate the dynamical nature of this term.
Of course, one must bear in mind that quantum chromodynamics,
as a quantum field theory, is an infinite-many body theory,
and that hadrons have an indefinite number of constituents.
To be sure, the derivation of Eq.~(\ref{eqn:density:energy}) did not require us to assume
that the hadron has a finite number of pointlike constituents,
but considering such an approximate scenario helps give intuition into the breakdown
of the energy density we have presented.

It is worth mentioning that the internal internal energy density is equivalent to
the light front energy density in the Drell-Yan frame in the formalism
of Wigner phase space distributions,
which is discussed in Ref.~\cite{Lorce:2018egm}.


\subsection{Stress tensor}

The light front stress tensor is given by
$T^{ij}_{\mathrm{LF}}(\mathbf{x}_\perp)$
with $i,j=1,2$.
As remarked previously in Ref.~\cite{Freese:2021czn},
the stress tensor contains contributions from hadron flow
and a ``comoving'' piece.
Accordingly, the stress tensor constitutes
a compound density in the manner of Eq.~(\ref{eqn:density:compound}).
We will now derive the exact form of this compound density.

The relevant matrix elements for the stress tensor are:
\begin{align}
  \frac{
    \langle P^+, \mathbf{p}'_\perp |
    T^{ij}(0)
    | P^+, \mathbf{p}_\perp \rangle
  }{2P^+}
  =
  \frac{P_\perp^i P_\perp^j}{P^+}
  A(t)
  +
  \frac{
    \Delta_\perp^i \Delta_\perp^j
    -
    \delta^{ij} \boldsymbol{\Delta}_\perp^2
  }{4P^+}
  D(t)
  \,,
\end{align}
for which the substitution rules of Eq.~(\ref{eqn:sub}) must be used.
The presence of two factors of $\mathbf{P}_\perp$ in front of $A(t)$
requires defining a new smearing function:
\begin{align}
  \label{eqn:smear:W:ij}
  \mathcal{W}^{ij}(\mathbf{R}_\perp)
  &=
  \int \mathrm{d}R^- \,
  \Psi^*(R^-,\mathbf{R}_\perp)
  \left(\frac{-i\olra{\boldsymbol{\nabla}}_\perp^i}{2}\right)
  \left(\frac{-i\olra{\boldsymbol{\nabla}}_\perp^j}{2}\right)
  \Psi(R^-,\mathbf{R}_\perp)
  \,,
\end{align}
which is related to the smearing function in Eq.~(\ref{eqn:smear:W:-}) by:
\begin{align}
  \mathcal{W}^{-}(\mathbf{R}_\perp)
  =
  \delta_{ij}
  \mathcal{W}^{ij}(\mathbf{R}_\perp)
  \,.
\end{align}
If we define the internal stress tensor as:
\begin{align}
  \label{eqn:stress:true}
  T^{ij}_{\mathrm{int.}}(\mathbf{b}_\perp)
  =
  \frac{1}{4}
  \int \frac{\mathrm{d}^2\boldsymbol{\Delta}_\perp}{(2\pi)^2}
  \Big(
    \Delta_\perp^i \Delta_\perp^j
    -
    \delta^{ij} \boldsymbol{\Delta}_\perp^2
    \Big)
    D(-\boldsymbol{\Delta}_\perp^2)
  \,
  e^{-i\boldsymbol{\Delta}_\perp\cdot\mathbf{b}_\perp}
  \,,
\end{align}
then the physical stress tensor can be written as a compound density:
\begin{align}
  \label{eqn:stress:physical}
  T^{ij}_{\mathrm{LF}}(\mathbf{x}_\perp)
  =
  \left\langle\frac{1}{P^+}\right\rangle
  \left(
  \int \mathrm{d}^2\mathbf{R}_\perp \,
  \mathcal{W}^{ij}(\mathbf{R}_\perp)
  T^{++}_{\mathrm{int.}}(\mathbf{x}_\perp-\mathbf{R}_\perp)
  +
  \int \mathrm{d}^2\mathbf{R}_\perp \,
  \mathcal{P}(\mathbf{R}_\perp)
  T^{ij}_{\mathrm{int.}}(\mathbf{x}_\perp-\mathbf{R}_\perp)
  \right)
  \,.
\end{align}
The quantity $T^{ij}_{\mathrm{int.}}(\mathbf{b}_\perp)$ is
(aside from a factor $P^+$)
identical to the pure stress tensor of Ref.~\cite{Freese:2021czn}.
The first and second terms in Eq.~(\ref{eqn:stress:physical})
can be identified as a flow tensor and comoving stress tensor
respectively~\cite{Fetter:1980str,Freese:2021czn},
so that the physical stress tensor essentially takes the classical
continuum form~\cite{Fetter:1980str}.
It is interesting to note that while diagonal elements of
$T^{ij}_{\mathrm{int.}}(\mathbf{b}_\perp)$ correspond to internal or static pressures
as seen by observers comoving with the motion of the hadron,
diagonal elements of $T^{ij}_{\mathrm{LF}}(\mathbf{x}_\perp)$
instead correspond to dynamic pressures,
which includes impulses that would be imparted by the hadron's motion
and which is akin to radiation pressure
(see discussion in Refs.~\cite{Binder1943:fluid,Fetter:1980str,Freese:2021czn,Freese:2022ibw}).


\subsection{Regarding wave packet localization}
\label{sec:fallacy}

Deriving Eq.~(\ref{eqn:density:energy}) for the energy density required
foregoing localization of the wave packet.
For a localized wave packet, the energy density actually diverges,
which can be understood in terms of the uncertainty principle~\cite{Miller:2018ybm}.

Let us consider what might happen if we attempt to localize the wave packet
by a procedure similar to that in
Refs.~\cite{Fleming:1974af,Freese:2021czn,Epelbaum:2022fjc,Panteleeva:2022khw},
but as in Ref.~\cite{Panteleeva:2022int} in particular, we absorb the divergence into
a normalization constant.
The wave function is localized by a scaling transformation:
\begin{align}
  \Psi(R^-,\mathbf{R}_\perp)
  \rightarrow
  \frac{1}{\sigma}
  \Psi\left(R^-,\frac{\mathbf{R}_\perp}{\sigma}\right)
  \,,
\end{align}
which transforms the smearing functions as:
\begin{subequations}
  \begin{align}
    \mathcal{P}(\mathbf{R}_\perp)
    & \rightarrow
    \frac{1}{\sigma^2}
    \mathcal{P}\left(\frac{\mathbf{R}_\perp}{\sigma}\right)
    \\
    \mathcal{W}^-(\mathbf{R}_\perp)
    & \rightarrow
    \frac{1}{\sigma^4}
    \mathcal{W}^-\left(\frac{\mathbf{R}_\perp}{\sigma}\right)
    \,.
  \end{align}
\end{subequations}
Using the variable change $\mathbf{Y}_\perp = \sigma\mathbf{R}_\perp$,
the energy density can be written:
\begin{align}
  T^{+-}_{\mathrm{LF}}(\mathbf{x}_\perp)
  &=
  \left\langle\frac{1}{2P^+}\right\rangle
  \left(
  \int \mathrm{d}^2\mathbf{Y}_\perp \,
  \mathcal{P}(\mathbf{Y}_\perp)
  T^{+-}_{\mathrm{int.}}(\mathbf{x}_\perp-\sigma\mathbf{Y}_\perp)
  +
  \frac{1}{\sigma^2}
  \int \mathrm{d}^2\mathbf{Y}_\perp \,
  \mathcal{W}^-(\mathbf{Y}_\perp)
  T^{++}_{\mathrm{int.}}(\mathbf{x}_\perp-\sigma\mathbf{Y}_\perp)
  \right)
  \,.
\end{align}
Defining the normalization:
\begin{align}
  N_\infty
  =
  \lim_{\sigma\rightarrow0}
  \frac{1}{\sigma^2}
  \left\langle\frac{1}{2P^+}\right\rangle
  \int \mathrm{d}^2\mathbf{Y}_\perp \,
  \mathcal{W}^-(\mathbf{Y}_\perp)
  \,,
\end{align}
the energy density in the limit $\sigma\rightarrow 0$ becomes:
\begin{align}
  T^{+-}_{\mathrm{LF}}(\mathbf{x}_\perp)
  \xrightarrow[\sigma\rightarrow0]{}
  N_\infty
  T^{++}_{\mathrm{int.}}(\mathbf{x}_\perp)
  \,.
\end{align}
The dynamical term was dropped because it remains finite in the
$\sigma\rightarrow0$ limit, and therefore is dominated by the kinetic term.
The problem should be immediately clear:
by localizing the compound density and keeping only the dominating term,
we lose important information about the internal structure of the hadron.
In fact, the internal density corresponding to dynamics is precisely what
was lost, and the remaining kinetic term is actually trivial and
uninteresting in this context.
On the other hand, organizing the physical density into a power series
in $\sigma$ will lead to additional complications,
since $\sigma$ appears within the internal densities.
In particular, derivatives of the internal $P^+$ density will mix with
the internal $P^-$ density at order $\sigma^0$,
so collecting order $\sigma^0$ terms together
as done in Refs.~\cite{Panteleeva:2022uii,Panteleeva:2023evj}
does not reproduce the internal $P^-$ density
as defined in our formalism.

This example should prove a cautionary tale about the localization procedure.
In several special cases---namely charge and $P^+$ density---the
internal light front densities were obtained from
localization because these are simple densities.
As we shall see in Sec.~\ref{sec:3D}, however,
none of the physical instant form densities
associated with the electromagnetic current or energy-momentum tensor are simple.
We are thus skeptical of the instant form densities that arise from wave packet localization.


\subsection{Numerical illustrations}

\subsubsection{Soft wall holographic pion}

The formulas we obtained above are fully general,
and do not require the application of light front dynamics.
Nonetheless, it may be especially enlightening to examine the densities
in a light front Hamiltonian model,
especially since having a light front Hamiltonian allows an explicit
breakdown of the energy density into kinetic and potential energy pieces.

As a simple example, we consider Brodsky's and de Teramond's model
of the pion in soft wall holographic QCD~\cite{Brodsky:2007hb}.
At large $Q^2$, or small transverse separations,
the system is approximately described by the light front Hamiltonian:
\begin{align}
  H_{\mathrm{internal}}
  =
  \frac{\mathbf{k}_\perp^2}{x}
  +
  \frac{\mathbf{k}_\perp^2}{1-x}
  +
  \kappa^4 x(1-x) y_\perp^2
  -
  2\kappa^2
  \,,
\end{align}
and the pion ground state is described
by the internal wave function:
\begin{align}
  \label{eqn:brodsky}
  \psi(x,\mathbf{y}_\perp)
  =
  \frac{\kappa}{\sqrt{\pi}}
  \sqrt{x(1-x)}
  e^{-\frac{1}{2} \kappa^2 x(1-x) \mathbf{y}_\perp^2}
  \,,
\end{align}
where $\kappa = 0.375$~GeV.
We have followed Soper~\cite{Soper:1976jc}
in denoting the transverse separation as $\mathbf{y}_\perp$,
and reserve $\mathbf{b}_\perp$ for the actual transverse distance from the
barycenter to the transverse coordinate---i.e., for the impact parameter.
The transverse separation is related to the actual transverse positions via:
\begin{align}
  \mathbf{y}_\perp
  =
  \left(\frac{\mathbf{r}_{1\perp} - \mathbf{R}_\perp}{1-x}\right)
  =
  \frac{\mathbf{b}_{1\perp}}{1-x}
  =
  -\left(\frac{\mathbf{r}_{2\perp} - \mathbf{R}_\perp}{x}\right)
  =
  - \frac{\mathbf{b}_{2\perp}}{x}
  \,.
\end{align}

\begin{figure}
  \includegraphics[width=0.49\textwidth]{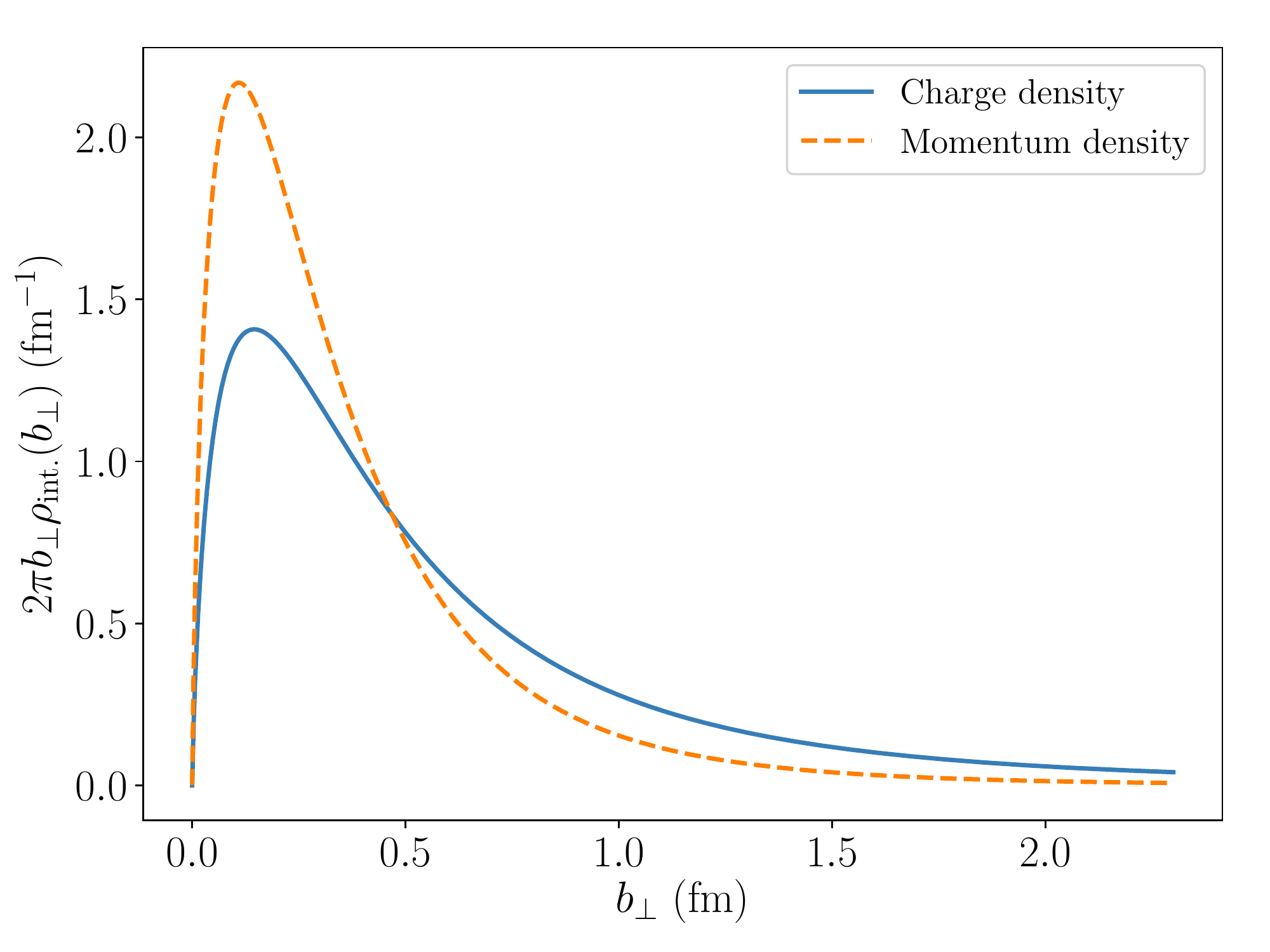}
  \includegraphics[width=0.49\textwidth]{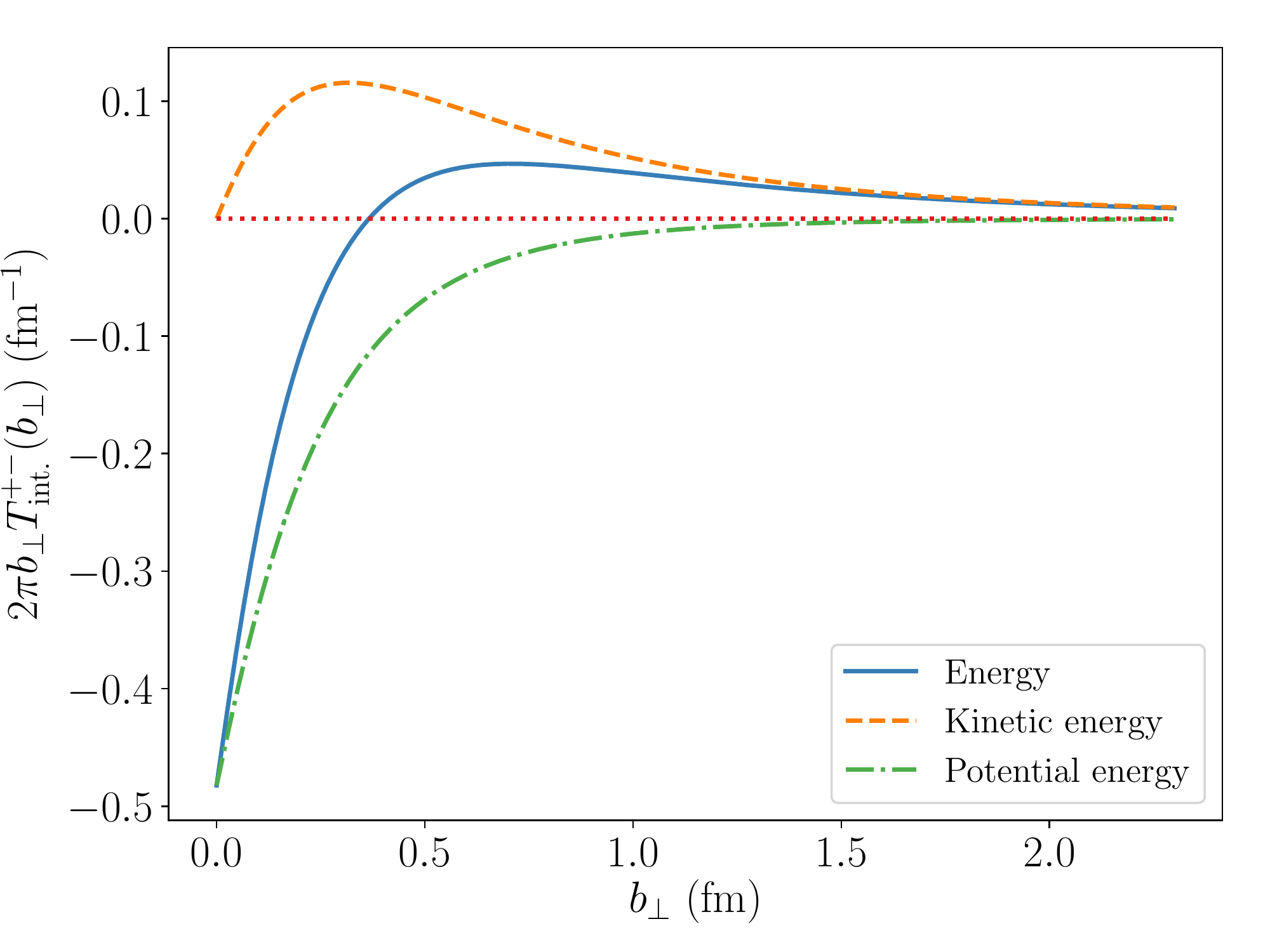}
  \caption{
    Internal pion densities in Brodsky's soft wall holographic model.
    (Left panel) charge and momentum densities.
    (Right panel) energy density, including its breakdown into
    kinetic and potential energy.
  }
  \label{fig:pion}
\end{figure}

The charge density is the most straightforward case to consider.
We assume a $\pi^+$ to avoid a trivially zero result.
There is an up quark with charge $Q_u$ and a down antiquark
with charge $-Q_d$, and the physical charge density is just given by the sum
of the single-particle distributions weighted by their charges:
\begin{align}
  j^+_{\mathrm{LF}}(\mathbf{x}_\perp)
  =
  \int \mathrm{d}R^-
  \int \mathrm{d}^2 \mathbf{R}_\perp
  \Big| \Psi(R^-,\mathbf{R}_\perp) \Big|^2
  \int_0^1 \mathrm{d}x
  \int \mathrm{d}^2\mathbf{y}_\perp
  \Big| \psi(x,\mathbf{y}_\perp) \Big|^2
  \Big\{
    Q_u \delta^{(2)}(\mathbf{x}_\perp - \mathbf{r}_{1\perp})
    -
    Q_d \delta^{(2)}(\mathbf{x}_\perp - \mathbf{r}_{2\perp})
    \Big\}
  \,,
\end{align}
where $Q_u = 2/3$ and $Q_d = -1/3$.
Comparing to Eq.~(\ref{eqn:j+:phys}) and Eq.~(\ref{eqn:j+:true}),
the internal internal charge density is:
\begin{align}
  j^+_{\mathrm{true}}(\mathbf{b}_\perp)
  =
  \int_0^1 \mathrm{d}x
  \left\{
    \frac{Q_u}{(1-x)^2}
    \left| \psi\left(x,\frac{\mathbf{b}_\perp}{1-x}\right) \right|^2
    -
    \frac{Q_d}{x^2}
    \left| \psi\left(x,\frac{\mathbf{b}_\perp}{x}\right) \right|^2
    \right\}
  \,.
\end{align}
The integral can be done analytically for the wave function in
Eq.~(\ref{eqn:brodsky}), and the result is:
\begin{align}
  j^+_{\mathrm{int.}}(\mathbf{b}_\perp)
  =
  \frac{\kappa^2}{\pi}
  e^{\kappa^2 b_\perp^2}
  \Big(
  E_1(\kappa^2 b_\perp^2)
  -
  E_2(\kappa^2 b_\perp^2)
  \Big)
  \,,
\end{align}
where
\begin{align}
  E_n(z)
  =
  \int_1^\infty \mathrm{d}t \,
  \frac{e^{-zt}}{t^n}
\end{align}
is the generalized exponential integral function~\cite{NIST:DLMF}.
The result is plotted in the left panel of Fig.~\ref{fig:pion}.

By a similar token, the internal $P^+$ density is given by:
\begin{align}
  T^{++}_{\mathrm{int.}}(\mathbf{b}_\perp)
  =
  \int_0^1 \mathrm{d}x
  \left\{
    \frac{x}{(1-x)^2}
    \left| \psi\left(x,\frac{\mathbf{b}_\perp}{1-x}\right) \right|^2
    +
    \frac{1-x}{x^2}
    \left| \psi\left(x,\frac{\mathbf{b}_\perp}{x}\right) \right|^2
    \right\}
  \,,
\end{align}
which can be evaluated using the wave function in Eq.~(\ref{eqn:brodsky}) as:
\begin{align}
  T^{++}_{\mathrm{int.}}(\mathbf{b}_\perp)
  =
  \frac{\kappa^2}{\pi}
  e^{\kappa^2 b_\perp^2}
  \Big(
  2
  E_1(\kappa^2 b_\perp^2)
  -
  4
  E_2(\kappa^2 b_\perp^2)
  +
  2
  E_3(\kappa^2 b_\perp^2)
  \Big)
  \,.
\end{align}
This result is also plotted in the left panel of Fig.~\ref{fig:pion}.

The last internal density we consider is the internal energy density.
Now, it is straightforward to find the part of the internal energy density
associated with quark kinetic energy,
using just the substitution rule
$\mathbf{k}_\perp \rightarrow -\frac{i}{2} \olra{\boldsymbol{\nabla}}_{(y)}$,
since this energy is spatially attached to the quark in question:
\begin{align}
  T^{+-}_{\mathrm{kin}}(\mathbf{b}_\perp)
  =
  -
  \int_0^1 \mathrm{d}x
  \left\{
    \frac{1}{x(1-x)^2}
    \psi^*\left(x,\frac{\mathbf{b}_\perp}{1-x}\right)
    \frac{\olra{\boldsymbol{\nabla}}^2}{4}
    \psi\left(x,\frac{\mathbf{b}_\perp}{1-x}\right)
    +
    \frac{1}{(1-x)x^2}
    \psi^*\left(x,\frac{\mathbf{b}_\perp}{x}\right)
    \frac{\olra{\boldsymbol{\nabla}}^2}{4}
    \psi\left(x,\frac{\mathbf{b}_\perp}{x}\right)
    \right\}
  \,.
\end{align}
For the wave function in Eq.~(\ref{eqn:brodsky}), we note:
\begin{align}
  - \frac{1}{4}
  \psi^*(x,\mathbf{y}_\perp)
  \olra{\boldsymbol{\nabla}}_i
  \olra{\boldsymbol{\nabla}}_j
  \psi(x,\mathbf{y}_\perp)
  =
  \frac{1}{2}
  \kappa^2 x(1-x)
  \delta_{ij}
  \psi^*(x,\mathbf{y}_\perp)
  \psi(x,\mathbf{y}_\perp)
  \,,
\end{align}
and the kinetic energy density can be explicitly found to be:
\begin{align}
  T^{+-}_{\mathrm{kin}}(\mathbf{b}_\perp)
  &=
  \frac{2\kappa^4}{\pi}
  e^{\kappa^2 b_\perp^2}
  \Big(
  E_2(\kappa^2 b_\perp^2)
  -
  E_3(\kappa^2 b_\perp^2)
  \Big)
  \,.
\end{align}
The potential energy cannot be assigned to either quark, however,
so some additional hypothesis about its spatial distribution is needed to proceed.
The $y_\perp^2$ term has the form of an elastic potential,
so we hypothesize that the corresponding energy is distributed along a string
or flux tube between the quarks.
The energy density operator so defined is thus:
\begin{align}
  \hat{T}^{+-}_{\mathrm{string}}(\mathbf{s}_\perp(\tau))
  =
  \lambda
  \int_0^1 \mathrm{d}\tau \, (\dot{\mathbf{s}}_\perp(\tau))^2
  \delta^{(2)}(\mathbf{x}_\perp - \mathbf{s}_\perp(\tau))
  \,,
\end{align}
where $\mathbf{s}_\perp(\tau)$ parametrizes the location of the string
in the transverse plane and $\lambda$ is the string energy density.
For a straight line,
\begin{align}
  \mathbf{s}_\perp(\tau)
  =
  \mathbf{r}_{2\perp}
  +
  \tau \mathbf{y}_\perp
  \,.
\end{align}
To reproduce the $y_\perp^2$ term,
an energy density $\lambda = \kappa^4 x(1-x)$ is sufficient.
For lack of a better hypothesis for the $-2\kappa^2$ term,
we distribute this energy evenly over the same string.
This makes the necessary string energy density:
\begin{align}
  \lambda(x,\mathbf{b}_\perp)
  =
  \kappa^4 x(1-x)
  -
  \frac{\kappa^2}{y_\perp^2}
  \,.
\end{align}
The internal energy density associated with this string is then:
\begin{align}
  T^{+-}_{\mathrm{pot}}(\mathbf{b}_\perp)
  =
  \int_0^1 \mathrm{d}x
  \int_0^1 \mathrm{d}\tau \,
  \frac{b_\perp^2}{(\tau-x)^4}
  \left| \psi\left(x,\frac{\mathbf{b}_\perp}{\tau-x}\right) \right|^2
  \lambda\left(x,\frac{\mathbf{b}_\perp}{\tau-x}\right)
  \,.
\end{align}
It is more difficult to obtain an exact analytic result for this integral
than for the other integrals we've considered.
An easy form to work with numerically for the result is:
\begin{align}
  T^{+-}_{\mathrm{pot}}(\mathbf{b}_\perp)
  =
  \frac{\kappa^4}{\pi}
  \left\{
    e^{\kappa^2 b_\perp^2}
    \Big(
    E_2(\kappa^2 b_\perp^2)
    -
    E_3(\kappa^2 b_\perp^2)
    \Big)
    -
    \frac{3 \sqrt{\pi}}{\kappa b_\perp}
    \int_0^{\pi/2} \mathrm{d}\theta \,
    \sin^2\theta \cos^2\theta
    \,
    \mathrm{erfc}(\kappa b_\perp \tan\theta)
    \right\}
  \,,
\end{align}
where $\mathrm{erfc}$ is the complementary error function~\cite{NIST:DLMF}:
\begin{align}
  \mathrm{erfc}(z)
  =
  \frac{2}{\sqrt{\pi}}
  \int_z^\infty \mathrm{d}z \,
  e^{-z^2}
  \,.
\end{align}

The results for the internal densities in this model are plotted in
Fig.~\ref{fig:pion}.
The left panel includes both the charge and momentum densities,
both of which are normalized to $1$.
The charge density is broader than the momentum density.
The right panel includes the internal energy density,
as well as its breakdown into potential and kinetic energy contributions.
An interesting feature of the energy density is that it can be negative,
in contrast to the momentum density which is positive-definite.

\subsubsection{Phenomenological multipole model}

Although illuminating, a light front Hamiltonian model is not necessary
to obtain the light front densities.
In fact, quantum chromodynamics is a quantum field theory and its bound
states thus have an indefinite number of constituents,
so any few-body model of hadron structure will necessarily be incomplete.
The definitions for light front densities do not require assuming the pion
to be a two- or even $N$-body system, however---they require only
the Poincar\'{e}-invariant form factors to be provided.

The form factors $A(t)$ and $D(t)$ can be obtained from phenomenology.
Multipole forms are reasonable for the form factors,
based on analyticity rules for the scattering amplitude.
For the pion in particular:
\begin{subequations}
  \begin{align}
    A(t)
    &=
    \frac{1}{1 - t/m_{f_2}^2}
    \\
    D(t)
    &=
    \frac{-1}{
      (1 - t/m_{f_2}^2)
      (1 - t/m_\sigma^2)
    }
    \,.
  \end{align}
\end{subequations}
The monopole form for $A(t)$ is supported by large-$N_c$
phenomenology~\cite{Masjuan:2012sk},
and the presence of an $f_2(1270)$ pole is suggested by spin-two meson
dominance as a gravitational analogy to vector meson dominance
in electromagnetic form factors.
Moreover, $m_{f_2} = 1270~\mathrm{MeV}$ reproduces the pion radius
reported by Kumano \textsl{et al.}'s analysis of Belle data~\cite{Kumano:2017lhr},
as well as the proton radius suggested by Kharzeev's analysis of
GlueX data~\cite{Kharzeev:2021qkd}
(using a dipole form for the proton's $A(t)$ form factor).
The presence of an additional $\sigma$ pole in $D(t)$ comes from dressing of
the graviton-quark vertex~\cite{Freese:2019bhb},
and the value $D(0) = -1$ comes from a
low-energy pion theorem~\cite{Novikov:1980fa,Voloshin:1980zf,Polyakov:2018zvc}.
Using $m_\sigma = 630~\mathrm{MeV}$ reproduces the $D(t)$ slope reported
in Ref.~\cite{Kumano:2017lhr}.

From these multipole forms of the form factors, we obtain
an internal momentum density:
\begin{align}
  T^{++}_{\mathrm{int.}}(\mathbf{b}_\perp)
  =
  \frac{m_{f_2}^2}{2\pi}
  K_0(m_{f_2} b_\perp)
  \,,
\end{align}
where $K_0(x)$ is a modified Bessel function of the second kind~\cite{NIST:DLMF}.
The internal energy density is given by:
\begin{align}
  T^{+-}_{\mathrm{int.}}(\mathbf{b}_\perp)
  =
  \left( m_\pi^2 - \frac{m_{f_2}^2}{4}\right)
  \frac{m_{f_2}^2}{2\pi}
  K_0(m_{f_2} b_\perp)
  +
  \frac{m_{f_2}^2}{4} \delta^{(2)}(\mathbf{b}_\perp)
  +
  \frac{1}{4\pi}
  \frac{ m_{f_2}^2 m_\sigma^2 }{ m_{f_2}^2 - m_\sigma^2 }
  \Big(
  m_{\sigma}^2 K_0(m_{\sigma} b_\perp)
  -
  m_{f_2}^2 K_0(m_{f_2} b_\perp)
  \Big)
  \,.
\end{align}

\begin{figure}
  \includegraphics[width=0.49\textwidth]{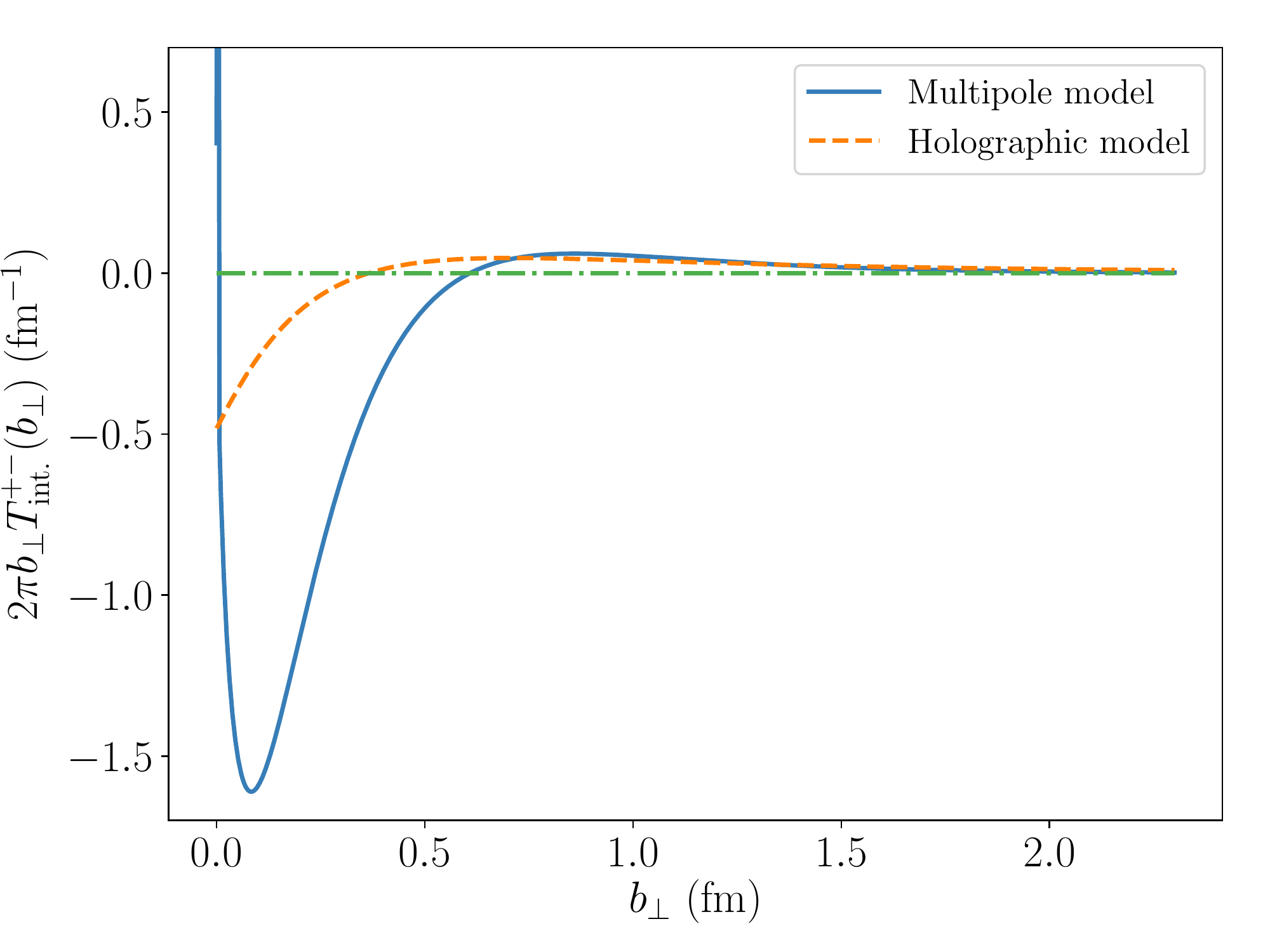}
  \includegraphics[width=0.49\textwidth]{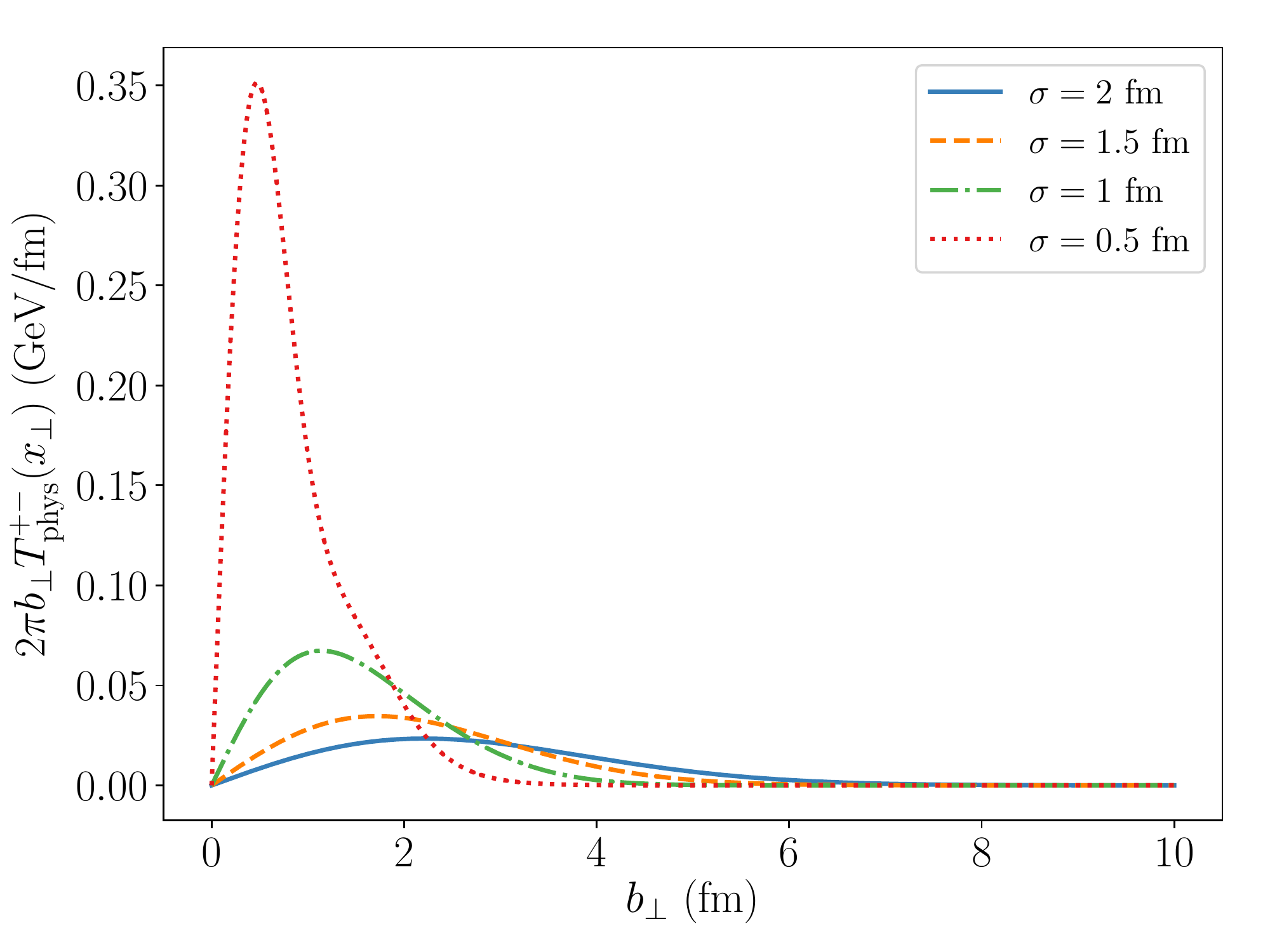}
  \caption{
    Pion energy densities associated with the phenomenological multipole model.
    (Left panel) compares the internal energy density in the multipole model
    to the soft wall holographic model.
    (Right panel) presents the \emph{physical} pion energy density
    (see Eq.~(\ref{eqn:density:energy}))
    for a Gaussian wave packet for several wave packet widths
    and with $2P^+ = m_\pi$.
  }
  \label{fig:compare}
\end{figure}

In the left panel of Fig.~\ref{fig:compare},
we compare the pion energy density obtained
with this phenomenological multipole parametrization to the result
found using the soft wall holographic model.
The energy densities look radically different,
suggesting a need for additional measurements of hard exclusive reactions
to further experimentally constrain the pion's gravitational form factors.
A curious feature that models have in common is that the internal energy density
becomes negative at short distances.

In the right panel of Fig.~\ref{fig:compare},
we present the \emph{physical} energy density using a Gaussian wave packet
of the form:
\begin{align}
  \Psi(\mathbf{R}_\perp)
  =
  \frac{1}{\sqrt{2\pi \sigma^2}}
  e^{-\frac{1}{4\sigma^2}\mathbf{R}_\perp^2}
  \,,
\end{align}
along with the phenomenological multipole model
at several values of the wave packet width $\sigma$
and with $2P^+ = m_\pi$ for concreteness.
It can be seen clearly that the physical density does not approach the
internal density as the wave packet is localized,
and is qualitatively quite different.
This is a stark manifestation of how wave packet localization will not
always distill the desired aspects of a hadron's internal structure.


\section{Internal densities in instant form coordinates}
\label{sec:3D}

To investigate physical densities in instant form coordinates in a wave packet
independent way, we need an analogous formula to that in
Eq.~(\ref{eqn:density:general}).
We shall proceed to obtain such a relation.
The procedure we follow here closely mirrors the derivation first
reported in Ref.~\cite{Li:2022ldb},
but with a different convention for defining the position-space wave function
(which in both our approach and in Li's is not a true wave function).

The Lorentz-invariant completeness relation for states in the Hilbert subspace
of single-particle spin-zero states is:
\begin{align}
  \int \frac{\mathrm{d}^3\mathbf{p}}{2E_{\mathbf{p}}(2\pi)^3}
  | \mathbf{p} \rangle \langle \mathbf{p} |
  =
  1
  \,.
\end{align}
We define the momentum space wave function as:
\begin{subequations}
  \begin{align}
    \Psi(\mathbf{p})
    =
    \frac{1}{\sqrt{2E_{\mathbf{p}}}}
    \langle \mathbf{p} | \Psi \rangle
    \,,
  \end{align}
  so that it obeys the normalization condition:
  \begin{align}
    \int \frac{\mathrm{d}^3\mathbf{p}}{(2\pi)^3}
    \big|
    \Psi(\mathbf{p})
    \big|^2
    =
    1
    \,.
  \end{align}
\end{subequations}
We similarly define the position-space wave function:
\begin{subequations}
  \begin{align}
    \Psi(\mathbf{r})
    =
    \int \frac{\mathrm{d}^3\mathbf{p}}{(2\pi)^3}
    \Psi(\mathbf{p})
    e^{i\mathbf{p}\cdot\mathbf{r}}
    \,,
  \end{align}
  so that it obeys the normalization condition:
  \begin{align}
    \int \mathrm{d}^3\mathbf{r} \,
    \big|
    \Psi(\mathbf{r})
    \big|^2
    =
    1
    \,.
  \end{align}
\end{subequations}
It should be noted that this position space wave function
is normalized differently than that of
Li \textsl{et al.}\ in Ref.~\cite{Li:2022ldb}.
Several of our results thus differ cosmetically from Li's,
but are in agreement when the difference in conventions
for the position space wave function is accounted for.
Following an analogous process to that outlined in Sec.~\ref{sec:lf},
it can be shown that:
\begin{align}
  \rho_{\mathrm{phys}}(\mathbf{x})
  =
  \int \mathrm{d}^3\mathbf{R}'_\perp
  \int \mathrm{d}^3\mathbf{R}_\perp
  \int \frac{\mathrm{d}^3\mathbf{P}}{(2\pi)^3}
  \int \frac{\mathrm{d}^3\boldsymbol{\Delta}}{(2\pi)^3}
  e^{i\mathbf{P}\cdot(\mathbf{R}'-\mathbf{R})}
  e^{-i\boldsymbol{\Delta} \cdot \left(\mathbf{x} - \frac{\mathbf{R}+\mathbf{R}'}{2}\right)}
  \Psi^*(\mathbf{R}')
  \frac{
    \langle \mathbf{p}' | \hat{O}(0) | \mathbf{p} \rangle
  }{
    2\sqrt{E_{\mathbf{p}} E_{\mathbf{p}'}}
  }
  \Psi(\mathbf{R})
  \,.
\end{align}
As with the light cone case, we need a substitution rule for $\mathbf{P}$
when it shows up in the matrix element.
The appropriate rule can be shown to be:
\begin{align}
  \label{eqn:sub:3D}
  \mathbf{P}
  \rightarrow
  -\frac{i}{2} \olra{\boldsymbol{\nabla}}
  \,.
\end{align}
Wherever $P^0$ shows up, it is necessary to write it out in terms of
$\mathbf{P}$:
\begin{align}
  P^0
  =
  \frac{E_{\mathbf{p}} + E_{\mathbf{p}'}}{2}
  =
  \frac{1}{2}\left(
  \sqrt{M^2 + \left(\mathbf{P}-\frac{\boldsymbol{\Delta}}{2}\right)^2}
  +
  \sqrt{M^2 + \left(\mathbf{P}+\frac{\boldsymbol{\Delta}}{2}\right)^2}
  \right)
  \,,
\end{align}
and then use Eq.~(\ref{eqn:sub:3D}) on each instance of $\mathbf{P}$ here.
With such substitutions implicitly applied,
the instant form equivalent of Eq.~(\ref{eqn:density:general}) is:
\begin{align}
  \rho_{\mathrm{phys}}(\mathbf{x})
  =
  \int \mathrm{d}^3\mathbf{R}_\perp
  \int \frac{\mathrm{d}^3\boldsymbol{\Delta}}{(2\pi)^3}
  e^{-i\boldsymbol{\Delta} \cdot (\mathbf{x} - \mathbf{R})}
  \Psi^*(\mathbf{R})
  \frac{
    \langle \mathbf{p}' | \hat{O}(0) | \mathbf{p} \rangle
  }{
    2\sqrt{E_{\mathbf{p}} E_{\mathbf{p}'}}
  }
  \Psi(\mathbf{R})
  \,,
\end{align}
where any $\mathbf{P}$ are to be implicitly understood in terms
of Eq.~(\ref{eqn:sub:3D}).
This is the furthest we can go without looking at specific local operators.
Obtaining a simple or finite compound density will depend on
the matrix element
factorizing in its $\boldsymbol{\Delta}$ and $\mathbf{P}$ dependence,
or else breaking down into a finite number of terms that factorize as such.
We will presently see that this does not happen.


\subsection{Electric charge and energy densities}

The appropriate matrix element for the electric charge density is:
\begin{align}
  \langle \mathbf{p}' | j^0(0) | \mathbf{p} \rangle
  =
  2 P^0 F(t)
  =
  \left(
  \sqrt{M^2 + \left(\mathbf{P}-\frac{\boldsymbol{\Delta}}{2}\right)^2}
  +
  \sqrt{M^2 + \left(\mathbf{P}+\frac{\boldsymbol{\Delta}}{2}\right)^2}
  \right)
  F(t)
  \,.
\end{align}
In instant form coordinates, $t$ is given by:
\begin{align}
  \label{eqn:t:3D}
  t
  =
  (p'-p)^2
  =
  -
  \boldsymbol{\Delta}^2
  +
  2\left(
  M^2
  +
  \mathbf{P}^2
  +
  \frac{\boldsymbol{\Delta}^2}{4}
  -
  \sqrt{
    \left(M^2 + \mathbf{P}^2 + \frac{1}{4}\boldsymbol{\Delta}^2\right)^2
    -
    (\mathbf{P}\cdot\boldsymbol{\Delta})^2
  }
  \right)
  \,.
\end{align}
The physical charge density thus takes the form:
\begin{multline}
  \label{eqn:j0:ugly}
  j^0_{\mathrm{phys}}(\mathbf{x})
  =
  \frac{1}{2}
  \int \mathrm{d}^3\mathbf{R}_\perp
  \int \frac{\mathrm{d}^3\boldsymbol{\Delta}}{(2\pi)^3}
  e^{-i\boldsymbol{\Delta} \cdot (\mathbf{x} - \mathbf{R})}
  \Psi^*(\mathbf{R})
  \frac{
    \left(
    \sqrt{M^2 + \frac{1}{4}\left(-i\olra{\boldsymbol{\nabla}}-\boldsymbol{\Delta}\right)^2}
    +
    \sqrt{M^2 + \frac{1}{4}\left(-i\olra{\boldsymbol{\nabla}}+\boldsymbol{\Delta}\right)^2}
    \right)
  }{
    \left[
      \left(M^2 - \frac{1}{4}\olra{\boldsymbol{\nabla}}^2 + \frac{1}{4}\boldsymbol{\Delta}^2\right)^2
      +
      \frac{1}{2} (\olra{\boldsymbol{\nabla}}\cdot\boldsymbol{\Delta})^2
      \right]^{1/4}
  }
  \\
  F\left(
  -
  \boldsymbol{\Delta}^2
  +
  2\left(
  M^2
  -
  \frac{1}{4} \olra{\boldsymbol{\nabla}}^2
  +
  \frac{1}{4} \boldsymbol{\Delta}^2
  -
  \sqrt{
    \left(M^2 - \frac{1}{4}\olra{\boldsymbol{\nabla}}^2 + \frac{1}{4}\boldsymbol{\Delta}^2\right)^2
    +
    \frac{1}{2} (\olra{\boldsymbol{\nabla}}\cdot\boldsymbol{\Delta})^2
  }
  \right)
  \right)
  \Psi(\mathbf{R})
  \,.
\end{multline}
The problem is immediately clear:
it is impossible to write this as a finite number of terms whose
$\mathbf{R}$ and $\boldsymbol{\Delta}$ dependence factorizes.
At best, we could expand the expression
as an infinite formal series in $\olra{\boldsymbol{\nabla}}$
and obtain an infinitely compound density.
Convergence of the series will depend on the wave packet taking a specific form,
which is precisely what we need to avoid to obtain a truly internal density.

The instant form energy density is similarly unfactorizable.
The relevant matrix element is~\cite{Polyakov:2018zvc}:
\begin{align}
  \langle \mathbf{p}' | T^{00}(0) | \mathbf{p} \rangle
  =
  \frac{1}{2}
  \left(
  \sqrt{M^2 + \left(\mathbf{P}-\frac{\boldsymbol{\Delta}}{2}\right)^2}
  +
  \sqrt{M^2 + \left(\mathbf{P}+\frac{\boldsymbol{\Delta}}{2}\right)^2}
  \right)
  ^2
  A(t)
  +
  \frac{\boldsymbol{\Delta}^2}{2}
  D(t)
  \,,
\end{align}
where $t$ is again defined in Eq.~(\ref{eqn:t:3D}).
This makes the physical instant form energy density:
\begin{align}
  T^{00}_{\mathrm{phys}}(\mathbf{x})
  =
  \int \mathrm{d}^3\mathbf{R}_\perp
  \int \frac{\mathrm{d}^3\boldsymbol{\Delta}}{(2\pi)^3}
  e^{-i\boldsymbol{\Delta} \cdot (\mathbf{x} - \mathbf{R})}
  \Psi^*(\mathbf{R})
  \left(
  \frac{(E_{\mathbf{p}} + E_{\mathbf{p}'})^2}{4\sqrt{E_{\mathbf{p}}E_{\mathbf{p}'}}}
  A(t)
  +
  \frac{\boldsymbol{\Delta}^2}{4\sqrt{E_{\mathbf{p}}E_{\mathbf{p}'}}}
  D(t)
  \right)
  \Psi(\mathbf{R})
  \,,
\end{align}
where we have neglected to explicitly write the substitution rules
in order to keep the formula compact.
However, a quick glance at Eq.~(\ref{eqn:j0:ugly}) should convince
the reader that this expression is not factorizable into either
a simple density in the manner of Eq.~(\ref{eqn:density:simple}),
nor a finite compound density in the manner of Eq.~(\ref{eqn:density:compound}).

In fact, we shall see below, that the instant form densities
are at best \emph{infinitely compound densities},
but convergence of this series depends on the form of the wave packet.


\subsection{Diffuse wave packets and Breit frame densities}
\label{sec:breit}

We have shown that
physical densities are infinitely compound
in instant form coordinates.
Nonetheless, it may be worth clarifying the approximate form of the
physical densities when they are expanded as a formal series in the derivative,
assuming that this expansion is controlled by a small parameter.
This can be the case for diffuse wave packets,
for which the derivative of the wave function is smaller than the wave function.

Let us consider the first few terms in the physical charge density
when expanded out in powers of the derivative.
The result is an infinite series of internal densities weighted by different smearing functions.
Li \textsl{et al.}~\cite{Li:2022ldb} refer to the resulting tower of internal densities as
multipole moment densities.
Odd powers of the derivative cancel in the expansion,
so we will consider the up to the second power.
The leading terms take the form:
\begin{subequations}
  \label{eqn:j0:phys}
  \begin{align}
    j^0_{\mathrm{phys}}(\mathbf{x})
    =
    \int \mathrm{d}^3\mathbf{R} \,
    \mathcal{P}(\mathbf{R})
    j^0_{\mathrm{Breit}}(\mathbf{x}-\mathbf{R})
    +
    \int \mathrm{d}^3\mathbf{R} \,
    \mathcal{Q}_{ij}(\mathbf{R})
    j^{ij}_{\mathrm{new}}(\mathbf{x}-\mathbf{R})
    +
    \ldots
    \,,
  \end{align}
  where the smearing functions:
  \begin{align}
    \mathcal{P}(\mathbf{R})
    &=
    \Psi^*(\mathbf{R})
    \Psi(\mathbf{R})
    \\
    \mathcal{Q}_{ij}(\mathbf{R})
    &=
    \Psi^*(\mathbf{R})
    \left( -\frac{i \olra{\boldsymbol{\nabla}}_i }{2M} \right)
    \left( -\frac{i \olra{\boldsymbol{\nabla}}_j }{2M} \right)
    \Psi(\mathbf{R})
  \end{align}
  and internal density functions:
  \begin{align}
    j^0_{\mathrm{Breit}}(\mathbf{b})
    &=
    \int \frac{\mathrm{d}^3\boldsymbol{\Delta}}{(2\pi)^3}
    F(-\boldsymbol{\Delta}^2)
    e^{-i\boldsymbol{\Delta} \cdot \mathbf{b}}
    \\
    j^{ij}_{\mathrm{new}}(\mathbf{b})
    &=
    \int \frac{\mathrm{d}^3\boldsymbol{\Delta}}{(2\pi)^3}
    \frac{
      \boldsymbol{\Delta}^i
      \boldsymbol{\Delta}^j
    }{
      1 + \frac{\boldsymbol{\Delta}^2}{4M^2}
    }
    \left(
    \frac{1}{8M^2}
    \frac{
      F(-\boldsymbol{\Delta}^2)
    }{
      1 + \frac{\boldsymbol{\Delta}^2}{4M^2}
    }
    +
    F'(-\boldsymbol{\Delta}^2)
    \right)
    e^{-i\boldsymbol{\Delta} \cdot \mathbf{b}}
  \end{align}
\end{subequations}
appear.
These of course are only the first two terms in an infinite series,
and we reiterate that the convergence of the series depends on
the wave packet being diffuse.
This series expansion for the physical charge density,
along with an explicit result for the ``new'' term,
was first reported by Li \textsl{et al.}\ in Ref.~\cite{Li:2022ldb}.
The apparently different result we obtain for the ``new'' term is due to
a difference in convention for defining the position-space wave function.

A similar expansion can be made for the instant form energy density:
\begin{align}
  \label{eqn:T00:phys}
  T^{00}_{\mathrm{phys}}(\mathbf{x})
  =
  \int \mathrm{d}^3\mathbf{R} \,
  \mathcal{P}(\mathbf{R})
  T^{00}_{\mathrm{Breit}}(\mathbf{x}-\mathbf{R})
  +
  \int \mathrm{d}^3\mathbf{R} \,
  \mathcal{Q}_{ij}(\mathbf{R})
  Q^{ij}_{\mathrm{new}}(\mathbf{x}-\mathbf{R})
  +
  \ldots
  \,,
\end{align}
where the standard Breit frame density is given by~\cite{Polyakov:2018zvc}:
\begin{align}
  T^{00}_{\mathrm{Breit}}(\mathbf{b})
  &=
  2M^2
  \int \frac{\mathrm{d}^3\boldsymbol{\Delta}}{(2\pi)^3}
  \frac{1}{\sqrt{4M^2 + \boldsymbol{\Delta}^2}}
  \left(
  A(-\boldsymbol{\Delta}^2)
  +
  \frac{\boldsymbol{\Delta}^2}{4M^2}
  \Big[ A(-\boldsymbol{\Delta}^2) + D(-\boldsymbol{\Delta}^2) \Big]
  \right)
  e^{-i\boldsymbol{\Delta} \cdot \mathbf{b}}
  \,,
\end{align}
and the new higher-order internal structure by:
\begin{multline}
  Q^{ij}_{\mathrm{new}}(\mathbf{b})
  =
  M^2
  \int \frac{\mathrm{d}^3\boldsymbol{\Delta}}{(2\pi)^3}
  \frac{1}{\sqrt{4M^2 + \boldsymbol{\Delta}^2}}
  \Bigg\{
    \delta^{ij}
    \left(
    A(-\boldsymbol{\Delta}^2)
    -
    \frac{ \boldsymbol{\Delta}^2 }{4M^2 + \boldsymbol{\Delta}^2}
    D(-\boldsymbol{\Delta}^2)
    \right)
    \\
    +
    2 \boldsymbol{\Delta}^i \boldsymbol{\Delta}^j
    \left(
    A'(-\boldsymbol{\Delta}^2)
    +
    \frac{ \boldsymbol{\Delta}^2 }{ 4M^2 + \boldsymbol{\Delta}^2 }
    \left[
      \frac{
        D(-\boldsymbol{\Delta}^2)
      }{
        4M^2 + \boldsymbol{\Delta}^2
      }
      +
      D'(-\boldsymbol{\Delta}^2)
      \right]
    \right)
    \Bigg\}
  e^{-i\boldsymbol{\Delta} \cdot \mathbf{b}}
  \,.
\end{multline}
For both the charge density and energy density,
the leading term for diffuse wave packets
is a convolution between the probability density of the barycenter
and the standard Breit frame density.
This suggests that, after all, the Breit frame densities do actually encode
an internal hadron density.
However, their interpretation as an internal density
relies on an assumption about the wave packet---namely, that it is spatially diffuse.

Through the consideration of diffuse wave packets,
we have reproduced the finding of Ref.~\cite{Li:2022ldb}
that Breit frame densities provide a leading-order description
for the internal densities of hadrons in diffuse wave packets.
Moreover,
by providing the next-to-leading order corrections,
we
have also
provided a means to numerically test the breakdown of
their applicability.
We shall numerically study this breakdown in Sec.~\ref{sec:breakdown}.


\subsection{Numerical illustration of Breit frame density breakdown}
\label{sec:breakdown}

We will illustrate the domain of applicability of Breit frame densities
using a simple wave packet---a Gaussian with average momentum $\mathbf{P}$:
\begin{align}
  \label{eqn:gauss}
  \Psi(\mathbf{R})
  =
  \frac{1}{(2\pi \sigma^2)^{3/4}}
  e^{-\frac{\mathbf{R}^2}{4\sigma^2}}
  e^{i\mathbf{P}\cdot\mathbf{R}}
  \,.
\end{align}
For such a wave packet:
\begin{align}
  -
  \frac{1}{4}
  \Psi^*(\mathbf{R})
  \olra{\boldsymbol{\nabla}}_i
  \olra{\boldsymbol{\nabla}}_j
  \Psi(\mathbf{R})
  =
  \left(
  \mathbf{P}_i
  \mathbf{P}_j
  +
  \frac{\delta_{ij}}{4\sigma^2}
  \right)
  \Psi^*(\mathbf{R})
  \Psi(\mathbf{R})
  \,,
\end{align}
meaning that the expansion considered in Sec.~\ref{sec:breit}
is valid when both $|\mathbf{P}|$ is small
(a slow wave packet)
and $\sigma$ is large (a diffuse wave packet).
The second-order smearing function for this wave packet is given by:
\begin{align}
  \mathcal{Q}_{ij}(\mathbf{R})
  &=
  \left(
  \frac{ \mathbf{P}_i \mathbf{P}_j }{M^2}
  +
  \frac{\delta_{ij}}{4\sigma^2M^2}
  \right)
  \mathcal{P}(\mathbf{R})
  \,.
\end{align}
We shall consider a case with proton-like kinematics,
using $M=940$~MeV and simple multipole models for the form factors:
\begin{subequations}
  \begin{align}
    F(-\boldsymbol{\Delta}^2)
    &=
    \frac{1}{
      (1 + \boldsymbol{\Delta}^2/m_\rho^2)^2
    }
    \\
    A(-\boldsymbol{\Delta}^2)
    &=
    \frac{1}{
      (1 + \boldsymbol{\Delta}^2/m_{f_2}^2)^2
    }
    \\
    D(-\boldsymbol{\Delta}^2)
    &=
    \frac{-1}{
      (1 + \boldsymbol{\Delta}^2/m_{f_2}^2)^3
    }
    \,,
  \end{align}
\end{subequations}
with $m_{\rho} = 776$~MeV
and $m_{f_2} = 1270$~MeV.
Although the proton is a spin-half particle,
this calculation is meant only for illustrative purposes.

\begin{figure}
  \includegraphics[width=0.49\textwidth]{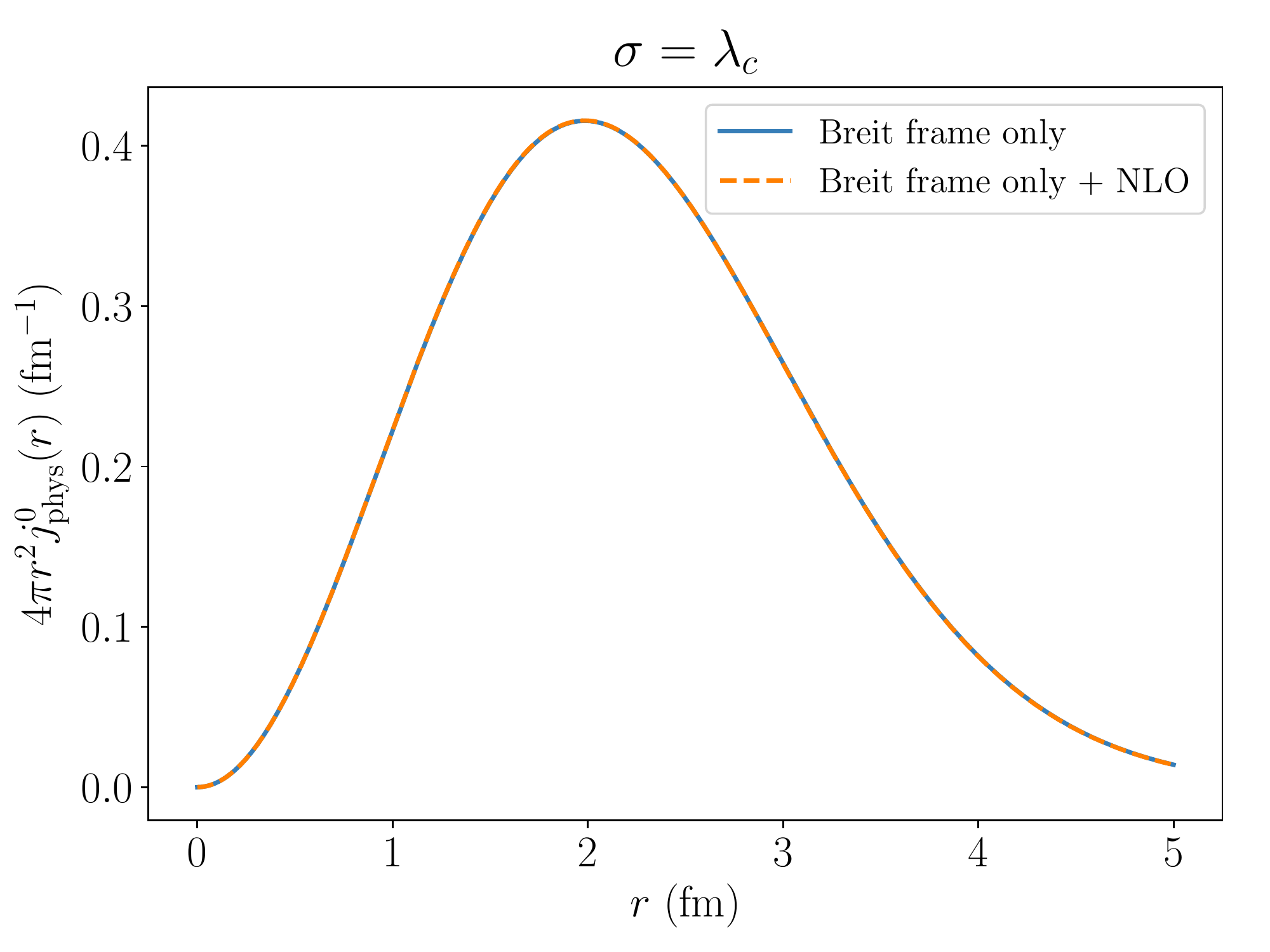}
  \includegraphics[width=0.49\textwidth]{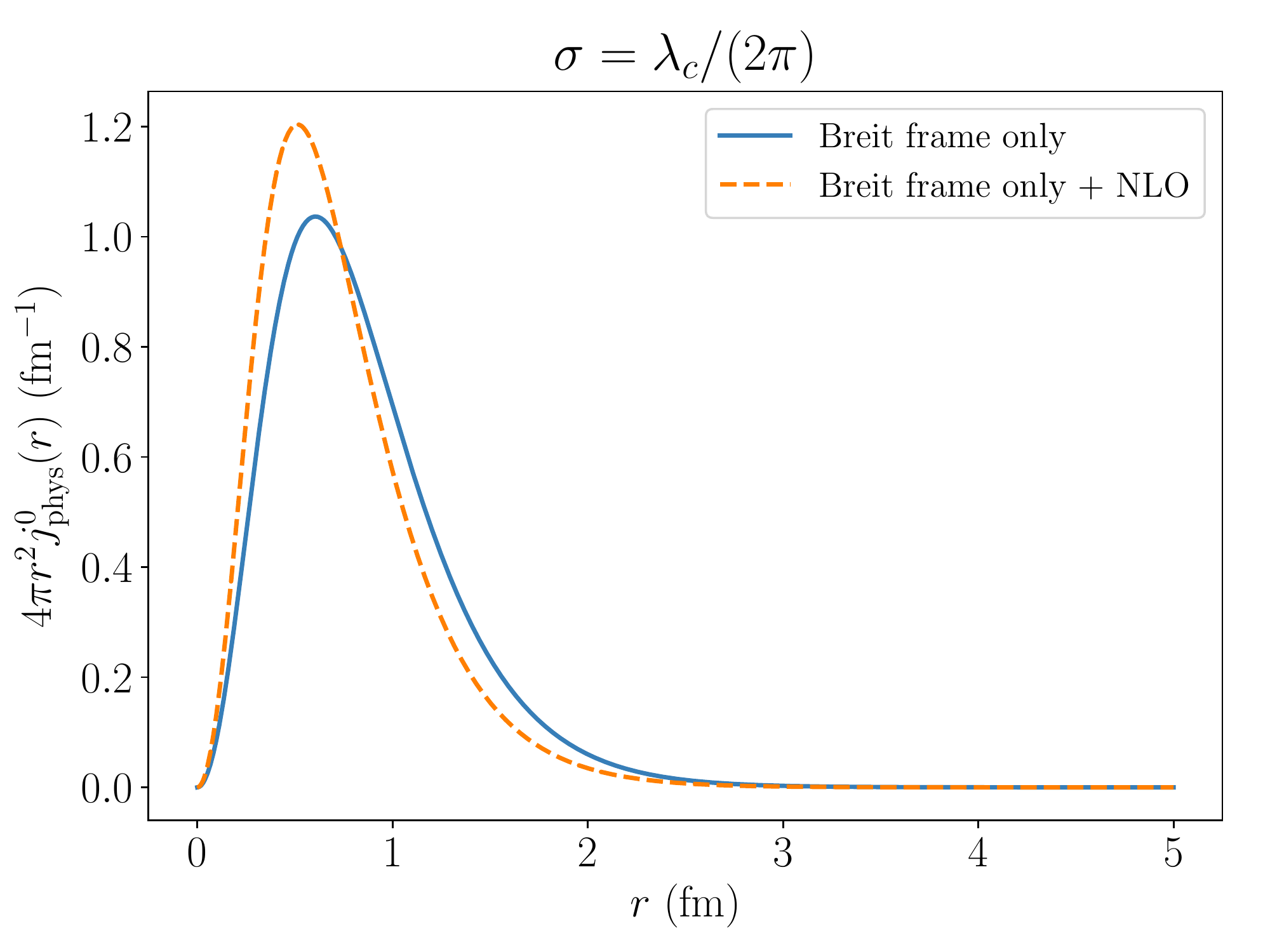}
  \caption{
    Leading approximations of the physical instant form charge density
    for a spin-zero target with proton-like properties,
    calculated according to Eq.~(\ref{eqn:j0:phys}).
    Both wave packets use Eq.~(\ref{eqn:gauss}) with $\mathbf{P}=0$.
  }
  \label{fig:j0:packets}
\end{figure}

For a numerical example, we consider wave packets with zero average momentum.
The leading approximations for the \emph{physical} instant form charge density
are presented in Fig.~\ref{fig:j0:packets} for two wave packet widths.
The more diffuse packet has a width equal to the Compton wave length:
\begin{align}
  \lambda_c
  =
  \frac{2\pi}{M}
  \approx
  1.3~\mathrm{fm}
  \,,
\end{align}
and the more localized packet has a width of the reduced Compton wavelength,
smaller by a factor $2\pi$.
For the more diffuse packet, corrections to the charge density from
structures beyond the Breit frame density are negligible.
For the more localized packet, however, there are significant corrections
to the physical density from the first higher-order term.
This indicates that the Breit frame charge density encodes one aspect of
the internal distribution of charge in the hadron, but does not tell the entire story,
since there are corrections when the hadron wave packet is localized to
smaller distance scales than its Compton wavelength.

\begin{figure}
  \includegraphics[width=0.49\textwidth]{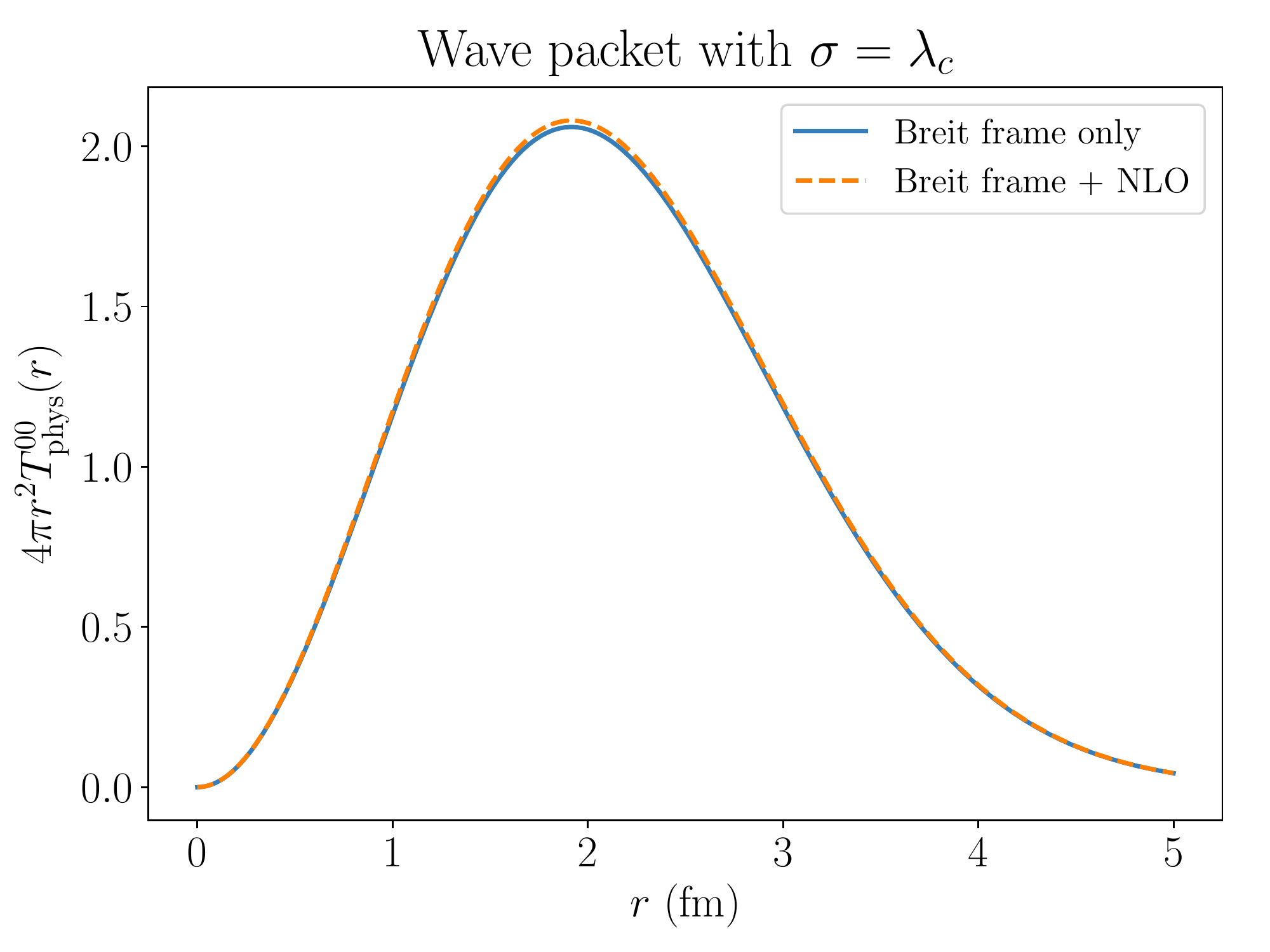}
  \includegraphics[width=0.49\textwidth]{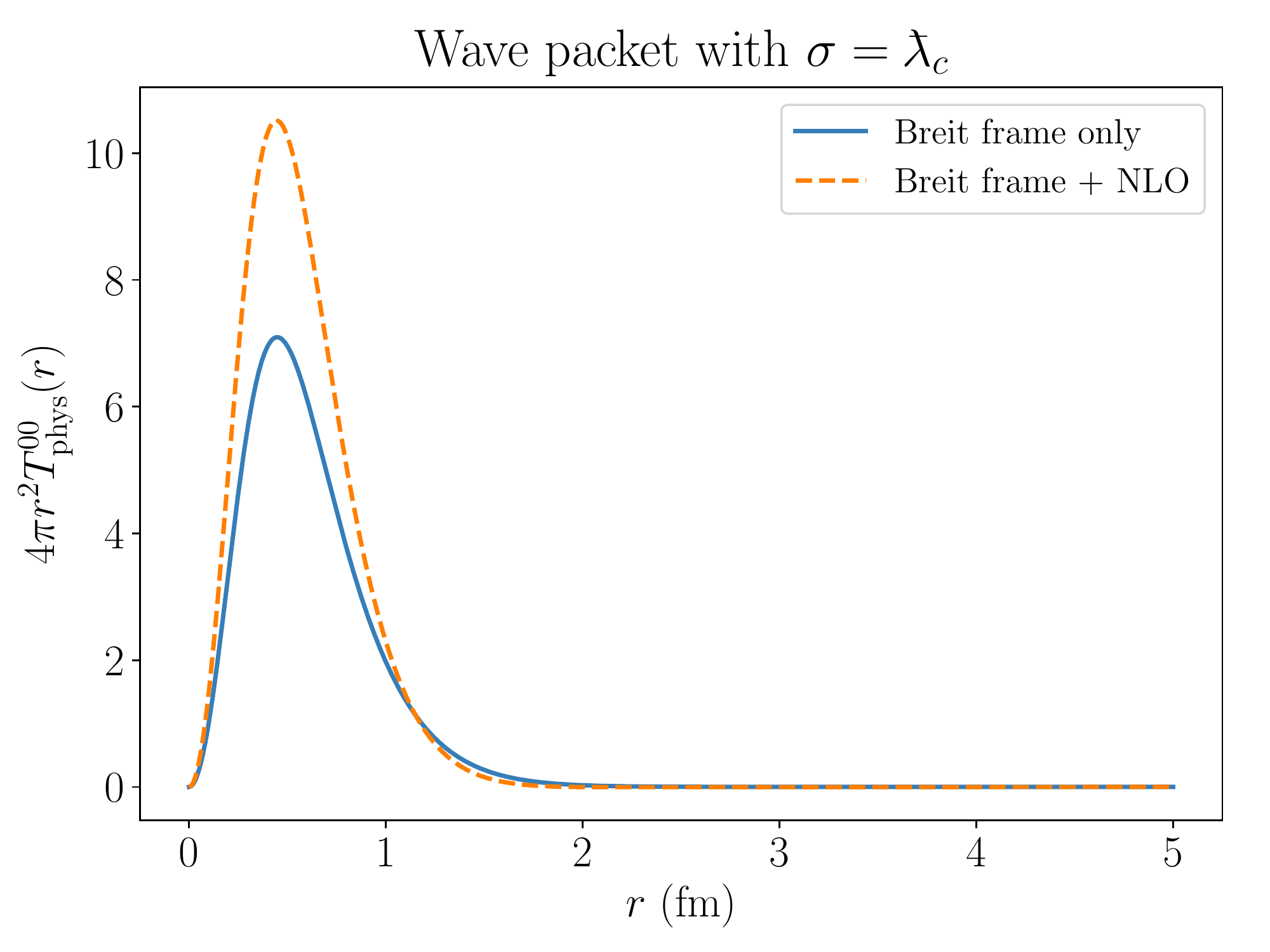}
  \caption{
    Leading approximations of the physical instant form energy density
    for a spin-zero target with proton-like properties,
    calculated according to Eq.~(\ref{eqn:T00:phys}).
    Both wave packets use Eq.~(\ref{eqn:gauss}) with $\mathbf{P}=0$.
  }
  \label{fig:T00:packets}
\end{figure}

We next repeat this exercise for the \emph{physical} instant form energy density
in Fig.~\ref{fig:T00:packets}.
Just as with the charge density, the Breit frame density dominates other
aspects of internal hadron structure for sufficiently diffuse wave packets,
and a width as large as the Compton wavelength is sufficient.
For smaller wave packets, however, the Breit frame density is no longer
sufficient to reproduce the physical density,
indicating that the Breit frame density
gives only a partial picture of a hadron's internal energy distribution.

There is one significant difference between Fig.~\ref{fig:j0:packets}
and Fig.~\ref{fig:T00:packets}:
the higher-order term in Fig.~\ref{fig:T00:packets} changes the normalization.
In fact, for the smaller wave packet, it should:
a more localized wave packet means a greater uncertainty in momentum,
and accordingly, a greater expectation value for the total energy.
The integral of the physical energy density should give the expected value
of the energy, but integrating the leading term in Eq.~(\ref{eqn:T00:phys})
only gives the mass.

It may be tempting to interpret the leading term of Eq.~(\ref{eqn:T00:phys})
as a mass density and the corrections as kinetic energy densities,
but this would be erroneous.
There is no ``kinetic charge'' in the charge density,
yet there were higher-order corrections to the instant form charge density.
The higher-order corrections to the energy density certainly contain
barycentric kinetic energy, but it is unclear how to separate these from
higher-order corrections to the mass density.
Isolating a barycentric kinetic energy for the light front energy density
(see Sec.~\ref{sec:lf})
was possible only because of the Galilean subgroup,
which does not manifest in instant form coordinates.


\section{Summary and conclusions}
\label{sec:conclusions}

In this work, we proposed and developed a formalism for identifying internal
densities of hadrons in a fully wave packet independent manner.
Physical densities are identified as matrix elements of local operators,
which necessarily mix internal structure with wave packet dependence.
When the dependence on internal structure and wave packet can be cleanly
separated within a convolution relation---or the physical density can be written
as a sum of such convolutions---an internal density can be defined.

When using light front coordinates with $x^-$ integrated out,
internal densities can always be identified,
owing to invariance of the remaining coordinates under the Galilean subgroup.
The scenario is thus analogous to the situation in
non-relativistic quantum mechanics.
For several simple cases such as the charge density and $P^+$ density,
the physical density approaches the internal density when the
wave packet is localized,
explaining why previous derivations of these quantities through
localized wave packets obtained the correct results.

By contrast, when using instant form coordinates, it is not possible
in general to separate wave packet and internal structure dependence.
A special case occurs when the hadron is prepared in a diffuse state
(i.e., broad in coordinate space),
in which case the physical density can be expanded as an infinite series
of convolutions between internal densities and
wave packet dependent smearing functions---a result
previously found by Li \textsl{et al.}~\cite{Li:2022ldb},
who refer to the tower of internal densities as ``multipole moment densities.''
For sufficiently broad wave packets,
on the order of the Compton wavelength or wider,
the physical density is dominated by a convolution between the
barycentric probability density and the conventional Breit frame density.
Thus, the Breit frame density can be identified as one of infinitely many
internal densities that describe hadron structure in instant form coordinates,
but one which dominates for diffuse wave packets.
Breit frame densities therefore have a legitimate claim
to describe \emph{an} aspect of hadron structure,
but they do not provide a complete description.

\begin{acknowledgments}
  We would like to thank
  Yang Li
  for illuminating discussions on the topics covered in this paper.
  This work was supported by the U.S.\ Department of Energy
  Office of Science, Office of Nuclear Physics under Award Number
  DE-FG02-97ER-41014.
\end{acknowledgments}




\bibliography{references.bib}

\end{document}